\documentclass[traditabstract,twocolumns]{aa}
\usepackage{graphicx}	% Including figure files
\usepackage{amsmath}	% Advanced maths commands
\usepackage{amssymb}	% Extra maths symbols
\usepackage{xcolor}     % Text color
\usepackage{soul}       % for strikeout, can be removed after comments are treated
\usepackage{array}
\usepackage{txfonts}
\usepackage{ulem}
\usepackage[colorlinks=true,linkcolor=blue,citecolor=blue, urlcolor=blue]{hyperref}
\bibpunct{(}{)}{;}{a}{}{,}

\usepackage{fontawesome}  % for GitHub icon
\usepackage{lipsum}
\usepackage{multirow}
\usepackage{colortbl}  % colored hlines in tables

\usepackage{rotating}
\usepackage{adjustbox}

\definecolor{fredcolor}{RGB}{204, 204, 0}
\definecolor{changesgreyblue}{RGB}{78, 129, 166}

\newcommand{\lcdm}{$\Lambda$CDM\xspace}
\newcommand{\herculens}{\textsc{Herculens}\xspace}
\newcommand{\lenstro}{\textsc{Lenstronomy}\xspace}
\newcommand{\starred}{\textsc{starred}\xspace}
\newcommand{\nifty}{\textsc{NIFTy}\xspace}
\newcommand{\jax}{\textsc{JAX}\xspace}
\newcommand{\ppxf}{\textsc{pPXF}\xspace}
\newcommand{\veldis}{\textsc{veldis}\xspace}
\newcommand{\elgor}{El Gordo\xspace}
\newcommand{\elanz}{El Anzuelo\xspace}

\newcommand{\zs}{\ensuremath{z_{\rm s}}\xspace}

\newcommand{\angs}{\ensuremath{\AA}\xspace}
\newcommand{\losvd}{\ensuremath{\sigma_{v,\rm los}}\xspace}
\newcommand{\thetaE}{\ensuremath{\theta_{\rm E}}\xspace}
\newcommand{\thetaESIS}{\ensuremath{\theta_{\rm E,SIS}}\xspace}

\newcommand{\Ds}{\ensuremath{D_{\rm s}}\xspace}
\newcommand{\Dds}{\ensuremath{D_{\rm ds}}\xspace}

\def\ks{${\, \mathrm{km}\, \mathrm{s}^{-1}}$\xspace}
\begin{document}

%-------------------------------------------------------------------

\title{\elgor needs \elanz: Probing the structure of cluster members with multi-band extended arcs in JWST data}

\titlerunning{JWST lens modeling of El Anzuelo}

\author{
A.~Galan\inst{\ref{tum},\ref{mpa}}\fnmsep\thanks{Corresponding author (\href{mailto:aymeric.galan@gmail.com}{aymeric.galan@gmail.com}).}, 
G.~B.~Caminha\inst{\ref{tum},\ref{mpa}},
J.~Knollmüller\inst{\ref{tum},\ref{origins},\ref{radboud}},
J.~Roth\inst{\ref{tum2},\ref{mpa},\ref{lmu}} 
and
S.~H.~Suyu\inst{\ref{tum},\ref{mpa}}
}

\institute{
Technical University of Munich, TUM School of Natural Sciences, Department of Physics, James-Franck-Str 1, 85748 Garching,
Germany \label{tum}
\goodbreak
\and
Max-Planck-Institut für Astrophysik, Karl-Schwarzschild-Str. 1, 85748 Garching, Germany \label{mpa}
\goodbreak
\and
ORIGINS Excellence Cluster, Boltzmannstr. 2, 85748 Garching, Germany \label{origins}
\goodbreak
\and
Radboud University, Heyendaalseweg 135, 6525 AJ Nijmegen, Netherlands \label{radboud}
\goodbreak
\and
Technische Universität M\"unchen (TUM), Boltzmannstr. 3, 85748 Garching, Germany \label{tum2}
\goodbreak
\and
Ludwig-Maximilians-Universität, Geschwister-Scholl-Platz 1, 80539 Munich, Germany \label{lmu}
}

\abstract{
Gravitational lensing by galaxy clusters involves hundreds of galaxies over a large redshift range and increases the likelihood of rare phenomena (supernovae, microlensing, dark substructures, etc.). Characterizing the mass and light distributions of foreground and background objects often requires a combination of high-resolution data and advanced modeling techniques. We present the detailed analysis of \elanz, a prominent quintuply imaged dusty star-forming galaxy ($\zs=2.29$), mainly lensed by three members of the massive galaxy cluster ACT-CL\,J0102$-$4915, also known as \elgor ($z_{\rm d}=0.87$). We leverage JWST/NIRCam images, which contain lensing features that were unseen in previous HST images, using a Bayesian, multi-wavelength, differentiable and GPU-accelerated modeling framework that combines \herculens (lens modeling) and \nifty (field model and inference) software packages. For one of the deflectors, we complement lensing constraints with stellar kinematics measured from VLT/MUSE data. In our lens model, we explicitly include the mass distribution of the cluster, locally corrected by a constant shear field. We find that the two main deflectors (L1 and L2) have logarithmic mass density slopes steeper than isothermal, with $\gamma_{\rm L1} = 2.23\pm0.05$ and $\gamma_{\rm L2} = 2.21\pm0.04$. We argue that such steep density profiles can arise due to tidally truncated mass distributions, which we probe thanks to the cluster lensing boost and the strong asymmetry of the lensing configuration. Moreover, our three-dimensional source model captures most of the surface brightness of the lensed galaxy, revealing a clump with a maximum diameter of $400$ parsecs at the source redshift, visible at wavelengths $\lambda_{\rm rest}\gtrsim0.6$ $\mu$m. Finally, we caution on using point-like features within extended arcs to constrain galaxy-scale lens models before securing them with extended arc modeling.
}

\keywords{Galaxies: clusters: general - Galaxies: clusters: individual: ACT-CL\,J0102$-$4915 - Galaxies: evolution - Infrared: galaxies - Gravitational lensing: strong - Methods: data analysis}

\maketitle
%-------------------------------------------------------------------

\section{Introduction \label{sec:intro}}

%\subsection{Strong gravitational lensing in galaxy clusters}

Extending over megaparsec scales and reaching up to quadrillion solar masses, massive galaxy clusters are direct tracers of the cosmic web and its evolution through cosmic time. Galaxy clusters emerge as information-rich targets for jointly studying rare phenomena such as microlensing and caustic-crossing events \citep[e.g.,][]{Dai2020,Williams2024}, distant supernovae \citep[e.g.,][]{Rodney2021,Frye2024}, and probing the cosmological evolution of the Universe \citep[e.g.,][]{Abbott2020,Bocquet2024}. The most massive clusters are made of several hundreds of co-evolving galaxies, showcasing in a single scene the many different stages of galaxy evolution predominantly in the redshift range $z\sim0 - 1$ \citep[e.g.,][]{Wen2012,Hilton2021,Klein2023,Bulbul2024}, probing the transition from a matter-dominated Universe to one driven by dark energy. Additionally, these clusters act as vast permeable screens standing between our telescopes and a plethora of more distant objects, enabling their detection up to redshifts $z\sim 13$ \citep[e.g.,][]{Adams2023,Wang2023_UNCOVER,Wang2024_UNCOVER} and probing the epoch of reionization. Galaxy clusters do not only magnify these distant objects but also distort and duplicate their images, which can take the form of giant arcs extending over several arcseconds \citep{Bayliss2011,Cava2018,Welch2022}. This striking phenomenon, called strong gravitational lensing, is a direct consequence of the extensive gravitational potential of foreground galaxy clusters and their alignment with populations of background sources. A notable property of strong gravitational lensing in clusters is the possible extreme time delays between the multiple images of time-varying sources \citep[which can reach several years, see e.g.,][]{Li2012}, which are used to geometrically measure absolute distances and cosmological parameters \citep[see e.g.,][]{Caminha2022,Kelly2023,Grillo2024,Pascale2024}, for which gains in precision are expected specifically with galaxy clusters \citep{Acebron2023,Bergamini2024}. 

Strong gravitational lensing in galaxy clusters is a natural and direct tool to characterize both visible and invisible massive structures of galaxies and their surroundings, which is key to deepen our understanding of their formation and evolution. The high multiplicity of lensed background sources can be used to constrain physical quantities describing individual galaxy members as well as their host dark matter halo through a process called lens modeling. On larger scales, typically out to the virial radius, strong lensing observations can effectively be combined with weak lensing features on the outskirts of clusters \citep[e.g.,][]{Bradac2006,Newman2013,Umetsu2018}, X-ray and Sunyaev-Zeldovich (SZ) emissions from hot ionized gas in the intra-cluster medium \citep[e.g.,][]{Newman2011,Planck_Ade2016}. However, at the scale of individual cluster members, challenges arise due to the large angular separation between the lensed images. These images often take the form of unresolved compact clumps within background sources that do not appear in the direct vicinity of cluster members. Such observational constraints necessitate specific modeling assumptions, such as parameterized mass profiles, to ensure that resulting lens models are both tractable and physically plausible \citep[as it is done in e.g., \textsc{lenstool},][]{JulloKneib2009}. Scaling relations, either based on observed luminosities or stellar kinematics, are also employed to intriduce priors or reduce the number of model parameters \citep[e.g.,][]{Bergamini2019,Caminha2023}. These models are generally sufficient to reproduce the observed lensing features and provide constraints on the fraction of the cluster mass that reside in individual galaxies, as well as measurements of the enclosed mass and dark matter fraction within these galaxies. With more lensing observables, such as extended arcs formed by resolved lensed sources, one can reasonably expect improvements in constraining the internal structure of cluster members, such as their density profiles over kiloparsec scales, and offsets between their baryonic and dark matter components \citep[][]{Schuldt2019,Wang2022}. As large-volume cosmological simulations improve in resolution---thus simulating more physical processes at smaller scales \citep[e.g.,][]{Schaye2015,Tremmel2017}---, it is key to constrain higher-order structural properties of cluster members to validate or invalidate predictions from different galaxy formation scenarios.

In particular, extended arcs that form around individual galaxies can be used to probe the radial mass density slope $\gamma$ measured at the Einstein radius $\theta_{\rm E}$. Different processes can impact $\gamma$, such as the cooling of baryonic matter followed by dark matter halo contraction \citep[e.g.,][]{Gnedin2004,Abadi2010}, itself compensated for by outflows from stellar winds or AGN feedback \citep[e.g.,][]{Johansson2012,Dubois2013}. Large elliptical galaxies, which are prevalent of cluster-scale and galaxy-scale strong lenses, were found to have a radial behavior remarkably close to an isothermal profile (i.e., $\gamma = 2$) from lensing and lens stellar dynamics constraints \citep[e.g.,][]{TreuKoopmans2002,Auger2010,Sonnenfeld2013}. This result has been termed the ``bulge-halo conspiracy'' in the literature \citep{Treu2006,DuttonTreu2014,Etherington2023_profiles}: despite the different radial behaviors of baryonic and dark matter profiles, their combination seems to lead to an isothermal density profile, with no evident signs for evolution with redshift \citep[e.g.,][]{Sonnenfeld2013,Etherington2023_profiles,Sahu2024_slope}. At the population level, this result has been recently confirmed using lensing-only constraints, larger samples, and more modern lens modeling techniques \citep{Shajib2021,Etherington2022_pyauto,Tan2023}.

%\subsection{The galaxy cluster \elgor and the galaxy-scale strong lens \elanz}

Significant improvements in lens models are achievable for a subset of cluster galaxies that coincidentally appear close in projection to extended lensed images of background sources, often visible as gravitational arcs. Alternatively, individual cluster members that are massive enough can create on their own additional strongly distorted and magnified extended arcs \citep[so-called galaxy-galaxy strong lensing events in cluters, e.g.,][]{Meneghetti2023}. These arcs are typically much fainter than the foreground lensing galaxies due to their larger distance, smaller intrinsic brightness and redshifted colors. Therefore, high signal-to-noise (S/N) and high resolution observations at wavelengths in which the lensed sources are brighter are necessary to provide useful constraints. In this regard, the \textit{James Webb} Space Telescope (JWST) and its imaging capabilities in near-infrared wavelengths, have provided the clearest observations of massive galaxy clusters with resolved extended sources. One striking example is the famous merging galaxy cluster ACT-CL\,J0102$-$4915 at redshift $z=0.87$, which has been one of the first clusters to be observed with JWST. Since its SZ-based discovery reported in \citet{Marriage2011}, several independent analyses confirmed its total mass of $\gtrsim10^{15}\ M_\odot$ giving it the nickname \elgor as it is the most massive known cluster at redshift $z\gtrsim0.8$, when the Universe was approximately 6 Gyr old \citep{Menanteau2012,Diego2023}. Recent images acquired using the Near Infrared Camera (NIRCam) on JWST as part of the Prime Extragalactic Areas for Reionization and Lensing Science (PEARLS) Guaranteed Time Observing (GTO) program \citep{Windhorst2023_PEARLS} have been publicly released in July 2023.

The current state of \elgor mass measurements (using the SZ effect, weak lensing shape measurements, or strong lensing) is given in Table~1 of \citet{Diego2023}. The first strong lensing analysis of the cluster have been achieved by \citet{Zitrin2013} with a light-traces-mass (LTM) model constrained by 9 families of multiple images deduced from \textit{Hubble} Space Telescope (HST) imaging data. Taking advantage of deeper HST images and thus more image families, \citet{Diego2020} relaxed most of the modeling assumptions from previous works by performing a free-form model of the cluster, obtaining an independent mass measurement. More recently, \citet{Caminha2023} leveraged deep spectroscopic data obtained on the Very Large Telescope (VLT) with the Multi Unit Spectroscopic Explorer (MUSE) to measure the redshifts of 23 multiply lensed galaxies and 167 cluster members, significantly refining the lens model compared to previous works. In our analysis, we make use of both their cluster-scale model their VLT/MUSE data. Two subsequent modeling analysis of \elgor have then been conducted using the JWST dataset from the PEARLS program. \citet{Diego2023} provided 28 new multiply imaged systems and obtained geometric redshifts (i.e., based on an initial lens model) for 37 systems, and then used the full set of 60 families to obtain a new free-form mass model. Shortly after, \citet{Frye2023} performed a new LTM model constrained by 56 multiply image systems to give further evidence of the two-component, cometary-like structure of the cluster, supporting that \elgor is undergoing a major merger event.

\begin{figure*}
    \centering
    \includegraphics[width=0.66\linewidth]{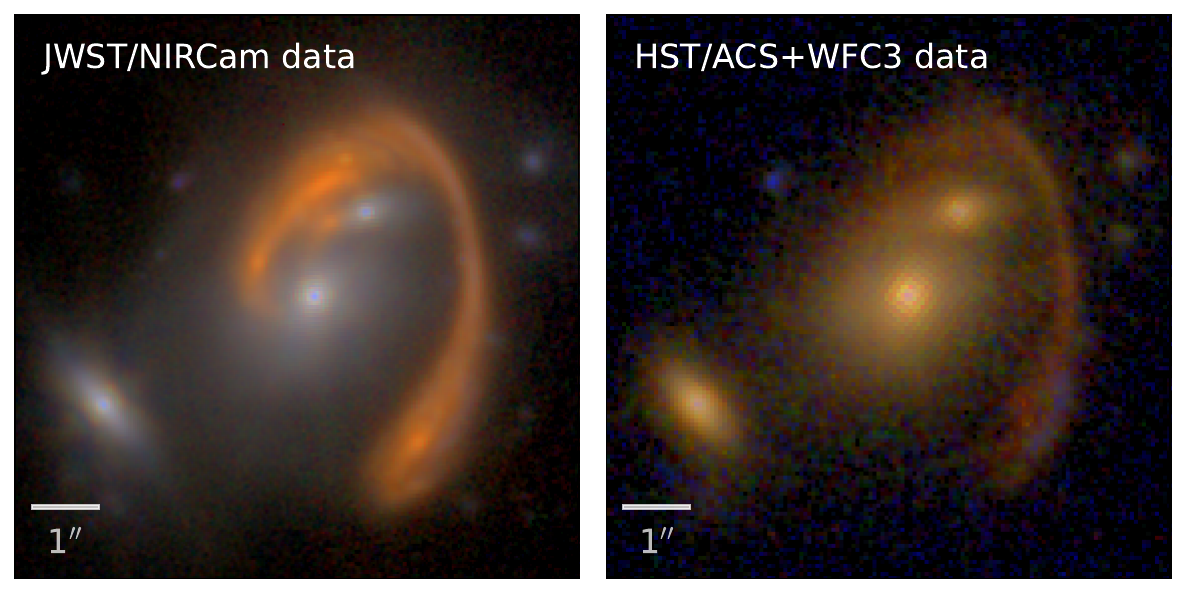}
    \includegraphics[width=\linewidth]{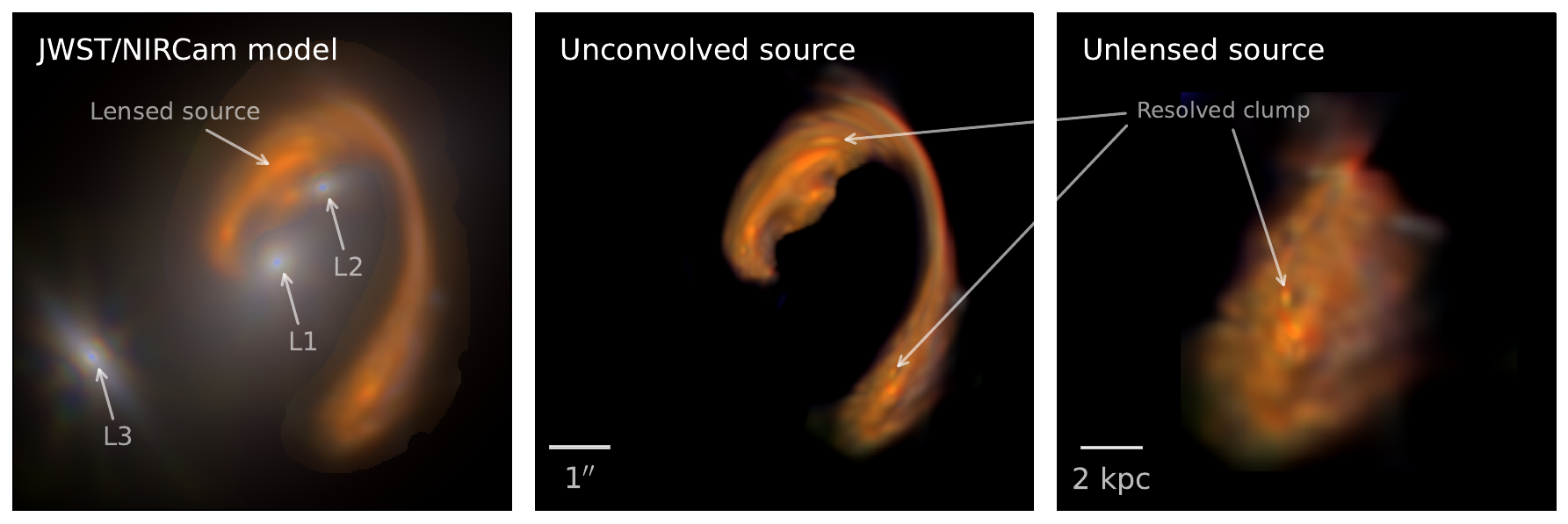}
    \caption{Color composite images of the dusty star-forming galaxy \elanz ($\zs\approx2.3$), strongly lensed by three members of the massive cluster \elgor (L1, L2, L3, $z_{\rm d}\approx0.9$). \textit{Top row, from left to right}: color composite built from JWST/NIRCam data used in this work (red: F444W; green: F410M, F356W; blue: F277W, F200W, F150W), archival HST/ACS+WFC3 data shown for comparison (red: F160W; green: F140W, F125W, F105W; blue: F775W, F625W, F606W, F435W). The exact weighting used for color composite images is given in Appendix~\ref{app:sec:color_composites}. \textit{Bottom row, from left to right}: best-fit model to the JWST/NIRCam data, unconvolved lensed source, unlensed source with arrows indicating the position of a clump revealed in our multi-band model.}
    \label{fig:model_color}
\end{figure*}

The JWST/NIRCam images of \elgor revealed features in lensed gravitational arcs that were unseen in previous HST images, including two remarkable ones that have been recently studied in more details. One is the giant arc La Flaca that extends over 20 arseconds, making it one of the longest known gravitational arcs to date. Although already visible in previous HST images, JWST images contain numerous new multiply imaged clumps within the arc, providing constraints on intervening low-mass perturbers in the mass range $10^9 - 10^{10}\ M_\odot$ \citep{Diego2023}. Another remarkable lensing feature of \elgor is called \elanz, that we show in the top left panel of Fig.~\ref{fig:model_color}. \elanz is a multiply imaged galaxy with a redshift of $z=2.29$ \citep[based on the most likely CO line detection at millimeters wavelengths,][]{Kamieneski2023} that appears as a highly distorted arc, bent around two cluster member galaxies (L1 and L2) and slightly influenced by a third one (L3). Previous HST images of \elanz, covering mainly optical wavelengths, feature only one clear distorted image of the background source as shown in the top right panel of Fig.~\ref{fig:model_color}. The better sensitivity and resolution of JWST/NIRCam images in the near-infrared reveal that the source is, in fact, multiply imaged at least two times (we find that it is multiply imaged five times). The lensed source, also known as DSFG\,J010249$-$491507, is a dusty star-forming galaxy whose multiple images have also been detected at radio wavelengths using the Atacama Large Millimeter/submillimeter Array \citep[ALMA, see][]{Cheng2023}. In a recent work, \citet{Kamieneski2023} analyzed in detail the morphology and photometric properties of \elanz using a combination of HST, JWST and ALMA datasets, and constructed a simple lens model based on compact multiply imaged features within the arcs. The authors measured the effective size of the lensed galaxy, characterized its dust distribution, and found that star formation is sensibly suppressed in its inner regions, supporting the hypothesis of ongoing inside-out quenching. We extensively compare our modeling results to the analysis of \citet{Kamieneski2023}.

%\subsection{This work: lens modeling of \elanz using a 3D correlated field}

The main goal of our work is to conduct the first detailed analysis of the mass distribution of individual \elgor members using \elanz strong lensing features. To this end we should employ a model that is able to capture the complexity of the lensed galaxy over multiple wavelengths, in order to provide stronger constraints on the lens model. We propose to use Gaussian processes, which have been employed in various applications (both within and outside astrophysics) as a mathematically simple yet very versatile model to capture complex correlations observed in data, for an arbitrary number of dimensions. The specific implementation of Gaussian processes we consider here is developed in the Bayesian framework of Information Field Theory \citep[IFT,][]{Ensslin2019} and is sometimes referred to as ``correlated fields''. Some recent astrophysical applications of such fields include the 4D (spatial, time and frequency) reconstruction of the M87 supermassive black hole \citep{Arras2022}, the 3D map of line-of-sight galactic magnetic field in the Milky Way \citep{Hutschenreuter2023}, or the 3D map of the dust distribution in the solar neighborhood \citep{Edenhofer2024}. These recent works have demonstrated that Gaussian processes within IFT can be used to model complex correlations seen in various types of observational data (spatial, temporal, frequency). These results, as well as the efficiency of the accompanying algorithms (low computation times despite high model flexibility), motivate us to apply a similar strategy to model the highly detailed surface brightness of the lensed dusty star-forming galaxy \elanz.

In our work, we perform pixel-level and multi-wavelength modeling of the extended arcs of \elanz as observed in the JWST/NIRCam dataset released by the PEARLS program. Building upon the work of \citet{Galan2022} that introduced the differentiable lens modeling code \herculens to describe complex mass and light distributions, we expand the code and use a 3D non-parametric correlated field model implemented in \nifty \citep{nifty1,nifty3,nifty5} to model the spatial and spectral morphology of \elanz source galaxy. Prior to lens modeling, we use the software package \starred \citep{Michalewicz2023} to reconstruct the complex point spread function (PSF) of the instrument. Our modeling pipeline is Bayesian, differentiable and runs on GPU, which significantly improves the computation time necessary to model the many data pixels. We also complement our lensing constraints using stellar kinematics measured on recent MUSE/VLT spectroscopic data using the template fitting software package \ppxf \citep{Cappellari2004,Cappellari2017}. To facilitate the access to our results and encourage future extensions of our analysis, we publicly release our models following the COOLEST lensing standard \citep{Galan2023_coolest}.

Two recent works employed lensed source models based on Gaussian processes. \citet{Karchev2022} used layers of Gaussian radial basis functions with predefined correlation lenghts (some of these layers being defined in image plane) to capture the multi-scale nature of highly complex simulated lensed sources. Very recently, \citet{Rustig2024_lenscharm} validated the lensing application of Gaussian processes and correlated fields within the IFT framework by modeling both simulated and real JWST/NIRCam observations of the strong lens system ${\rm SPT}0418-47$. Unlike in our work, \citet{Rustig2024_lenscharm} used a Matérn covariance kernel (in the Fourier domain) for their Gaussian process source model, reserving the use of the non-parametric correlated field to capture complexity in the lens mass distribution.

This paper is organized as follows. In Section~\ref{sec:data} we describe the imaging and spectroscopic datasets used in this work. In Section~\ref{sec:kinematics}, we present stellar kinematics measurements of a subset of the deflectors of \elanz, used as a prior in the lens modeling analysis described in Section~\ref{sec:image_modeling}. The resulting models are presented in Section~\ref{sec:results}, from which we select a subset of key results that are discussed further in Section~\ref{sec:discussion}. Finally, we summarize and conclude our work in Section~\ref{sec:conclusion}.

Throughout this work, we assume a Lambda cold dark matter (\lcdm) cosmology with $H_0 = 70\ {\rm km}\,{\rm s}^{-1}\,{\rm Mpc}^{-1}$ and $\Omega_{\rm m} = 0.3$. This cosmology leads to angular sizes of approximately $7.7\ {\rm kpc\, arcsec}^{-1}$ at redshift $z=0.87$ and $8.2\ {\rm kpc\, arcsec}^{-1}$ at redshift $z=2.29$.

\section{Data sets \label{sec:data}}

\subsection{Imaging data}

We primarily use imaging data obtained as part of the PEARLS program (ID: 1176, PI: Windhorst), in particular observations of the cluster ACT-CL\,J0102$-$4915 (\elgor) that were publicly released in July 2023 \citep{Windhorst2023_PEARLS}. The \elgor cluster has been selected for PEARLS because of its extreme mass ($M\sim 10^{15}\ M_\odot$), its elongation due to a double-peak post-collision morphology, a large number of lensed sources, and low contamination by intra-cluster light. The PEARLS data set for \elgor consists in NIRCam images in 8 filters---F090W, F115W, F150W, F200W, F277W, F356W, F410M, F444W---covering a wavelength range $0.9\,\mu{\rm m} - 4.5\,\mu{\rm m}$. The acquisitions were all taken on July 29, 2022, and the images were calibrated and reduced as described in great detail in \citet{Windhorst2023_PEARLS}. In particular, each individual frame was aligned to the Gaia DR3 reference frame \citep{Gaia2023}, such that images in each NIRCam filter are sufficiently aligned for our purposes. We use the FITS files with suffix ``\texttt{\_20230718}'' available on the PEARLS website\footnote{PEARLS website: \url{https://sites.google.com/view/jwstpearls}.}, which include the drizzled science exposures (``\texttt{\_drz}'') and associated weight maps (``\texttt{\_wht}''). All drizzled exposures have a pixel size of $0\farcs03$, and North is aligned with the vertical. We do not use the publicly released data products available on the Mikulski Archive for Space Telescopes, as we notice issues with background subtraction and severe residual noise patterns, which were almost entirely corrected for by the PEARLS reduction \citep[e.g., with careful treatment of $1/f$ noise patterns;][]{Windhorst2023_PEARLS}. The full width at half maximum (FWHM) of the point spread function (PSF) of the NIRCam images of \elgor is in the range $0\farcs067 - 0\farcs171$ \citep[see Table 1 of][]{Windhorst2023_PEARLS}. The first panel of Fig.~\ref{fig:model_color} shows a color composite image using 6 of the filters from the NIRCam data set.

Although to a lesser extent, we also use archival HST/ACS+WFC3 data of \elgor from the Reionization Lensing Cluster Survey \citep[RELICS, ID 14096, PI:][]{Coe2019}. We use this data mainly as a check to ensure alignment between (1) the cluster-scale model of \cite{Caminha2023} based on HST, and (2) the JWST/NIRCam data set. We used subset of the filters (drizzled to $0\farcs06$ pixel size) in the top right of  Fig.~\ref{fig:model_color}.

\subsection{Spectroscopic data}

\begin{figure*}
    \centering
    \includegraphics[width=\linewidth]{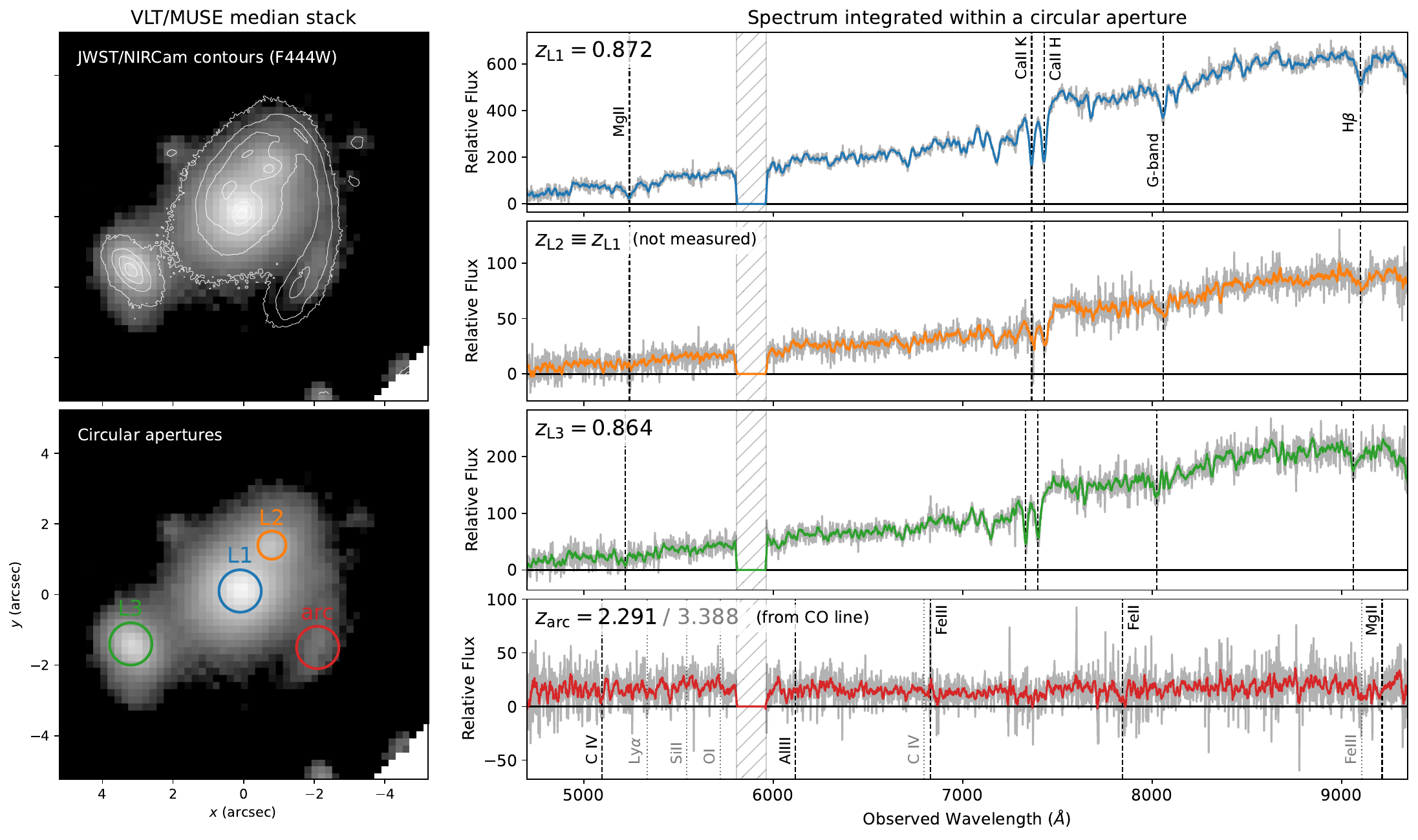}
    \caption{Extraction of MUSE aperture spectra within circular apertures, for L1, L2, L3 and the western part of \elanz arc. Top left panel: luminosity-weighted median stack of the MUSE datacube, with contours from NIRCam/F444W observations. Bottom left panel: MUSE median stack with circular apartures used to extract spectra. Circular apertures have a radius of $0\farcs6$ for L1, L3 and the arc, and $0\farcs3$ for L2. Right panels: corresponding 1D spectra (in gray) in the observed frame, with absorptions lines indicated, as well as smoothed spectra (in color). The hatched region indicates the gap caused to laser guiding. The estimated redshift of each component is also indicated in the upper left corner. We measure the redshifts for L1 and L3, but assume that L2 is at the same redshift as L1 and that the source is at redshift $z=2.291$ \citep[][using a CO line detection in ALMA observations]{Kamieneski2023}. Some key absorption lines are indicated by thin dashed gray lines (line labels are indicated in top and bottom panels only to avoid clutter). For the arc, we indicate two sets of lines corresponding to our assumed redshift (black and dashed lines) and the alternative possible redshift $z=3.388$ \citep[gray and dotted lines, for details see][]{Kamieneski2023}. In this work we only use L1 and L3 spectra to extract their stellar kinematics measurements, as L2 is highly blended with L1 and the signal from the arc is too faint.}
    \label{fig:muse_spectra}
\end{figure*}

We complement JWST imaging data with integral field unit spectroscopic data from the MUSE instrument on the VLT. We use data obtained between December 2018 and September 2019 under the ESO program ID 0102.A-0266 (PI: Caminha), which has been reduced and used in \citet{Caminha2023} to measure redshifts of objects within the field of \elgor. The final datacube covers the wavelength range $4700\,\angs - 9350\,\angs$ (with a gap between $5805\,\angs$ and $5965\,\angs$ due to laser guiding). The full field-of-view covers an area of $\sim 3\ {\rm arcmin}^2$, although here we only use a small cutout centered on \elanz. The FWHM of these MUSE observations falls in the range $0\farcs55 - 0\farcs60$ \citep{Caminha2023}. The median average of the datacube centered on \elanz and spectra for each deflectors are shown in Fig.~\ref{fig:muse_spectra}.

\section{Stellar kinematics measurements \label{sec:kinematics}}

\subsection{Redshift and velocity dispersion measurements \label{ssec:redshift_vel_disp}}

We measure redshifts and line-of-sight (LOS) central velocity dispersions from the MUSE observations of \citet{Caminha2023}. We first extract integrated spectra for the three deflectors within circular apertures as shown on Fig.~\ref{fig:muse_spectra}. The aperture radius is set to $0\farcs6$ which approximately corresponds to the MUSE resolution. For comparison, we also show one dimensional spectra corresponding to the brightest part of the arc, as well as a tentative extraction for L2. For the latter we use an aperture twice as small in order to reduce contamination by L1. We note that we only attempt to measure stellar kinematics of L1 and L3 for which we can detect clear absorption features. The spectrum of L2 suggests that this galaxy lies at a redshift very close to L1. However, extracting robust stellar kinematics properties for L2 would require to properly separate the contributions from L1 and L2, which is outside the scope of this work.

Due to low signal-to-noise (S/N), we are unable to detect reliable features in the spectrum of the lensed arc, as shown in the bottom right panel of Fig.~\ref{fig:muse_spectra}, such that we cannot robustly measure its redshift by fitting stellar templates. However, we indicate in Fig.~\ref{fig:muse_spectra} some absorption lines assuming the most probably redshift $z=2.291$ inferred by \citet{Kamieneski2023} assuming the CO line detection in ALMA data corresponds to the CO(3-2) transition. If instead it corresponds to the CO(4-3) transition, \citet{Kamieneski2023} mention that the redshift would change to $z=3.388$. For comparison, we indicate the spectral lines corresponding to this alternate redshift in Fig.~\ref{fig:muse_spectra} with dotted gray lines (bottom right panel). While the low S/N of the MUSE spectrum does not provide a definitive answer, the indicated absorption lines seem to align better with possible spectral features at $z=2.291$ compared to $z=3.388$. We note that the difference in angular-to-physical size between these two redshift estimates is only $0.8\,{\rm kpc}\,{\rm arcsec}^{-1}$. In the remaining of the work, we follow \citet{Kamieneski2023} and assume $\zs=2.291$ for the source redshift, and we show in Appendix~\ref{app:sec:alt_src_redshift} that the alternative value for $\zs$ impacts only marginally our results.

We use the spectral template fitting software package \ppxf \citep{Cappellari2004,Cappellari2017} to model L1 and L3 spectra to measure their line-of-sight central velocity dispersion \losvd and redshift. As the S/N of the spectra are significantly lower below 6000 \angs, we consider only wavelength interval $6650\,\angs - 9300\,\angs$ for all kinematics measurements. We run \ppxf multiple times in order to marginalize over possible systematic effects due to specific modeling assumptions. First, we vary the orders of the additive polynomial that captures large-scale variations in the continuum, from no polynomial up to order 8 polynomials. Second, we repeat the measurements (with all polynomial orders) varying this time the stellar template library. We use the Indo-US library \citep{Valdes2004_IndoUS}, which has a spectral resolution (FWHM) of $\sigma_{\rm temp, indo}=0.57$ \angs (for a dispersion of $0.4$ \angs), and the MILES library \citep{SanchezBlazquez2006_MILES}, which has a spectral resolution (FWHM) of $\sigma_{\rm temp, miles}=1.07$ \angs (for a dispersion of $0.9$ \angs). In empty patches in the vicinity of the lens, we fit Gaussian profiles to 7 sky emission lines to estimate the line spread function (LSF) and find $\sigma_{\rm lsf, obs} = 1.34$ \angs (for a dispersion of $1.25$ \angs). Assuming a redshift of $z=0.87$, this leads to a rest-frame value of $\sigma_{\rm lsf, rest} = 0.72$ \angs, which we use as the spectral resolution of the MUSE observations. As the Indo-US library has a better resolution as our spectra, we need to convolve the spectral templates with a kernel of squared width $\sigma^2_{\rm diff} = \sigma^2_{\rm lsf, rest} - \sigma^2_{\rm temp, indo}$ and run the optimization to obtain the best-fit \losvd value. We apply a similar treatment for measurements obtained with the MILES library; however, these templates have a slightly lower resolution compared to our spectra, thus we apply the correction after the \ppxf fit in order to retrieve the corrected best-fit \losvd value \citep[see Eq. 5 of][]{Cappellari2017}. Since \ppxf requires an estimation of the noise per wavelength, and the variance from the MUSE data cube is unreliable \citep[see e.g., Sect.~3.1.5 of][]{Bacon2017}, we follow \ppxf examples\footnote{\url{https://github.com/micappe/ppxf_examples}} and assume a uniform error equal to unity throughout the entire wavelength range. We emphasize that this choice of noise level does not impact the best-fit values, but only error estimates based directly on the $\chi^2$, which is re-estimated after the fit. To perform all the aforementioned processing steps and measurements, we use an extended version of the \ppxf wrapper code \veldis \citep{Mozumdar2023}.

\begin{figure*}
    \centering
    \includegraphics[width=\linewidth]{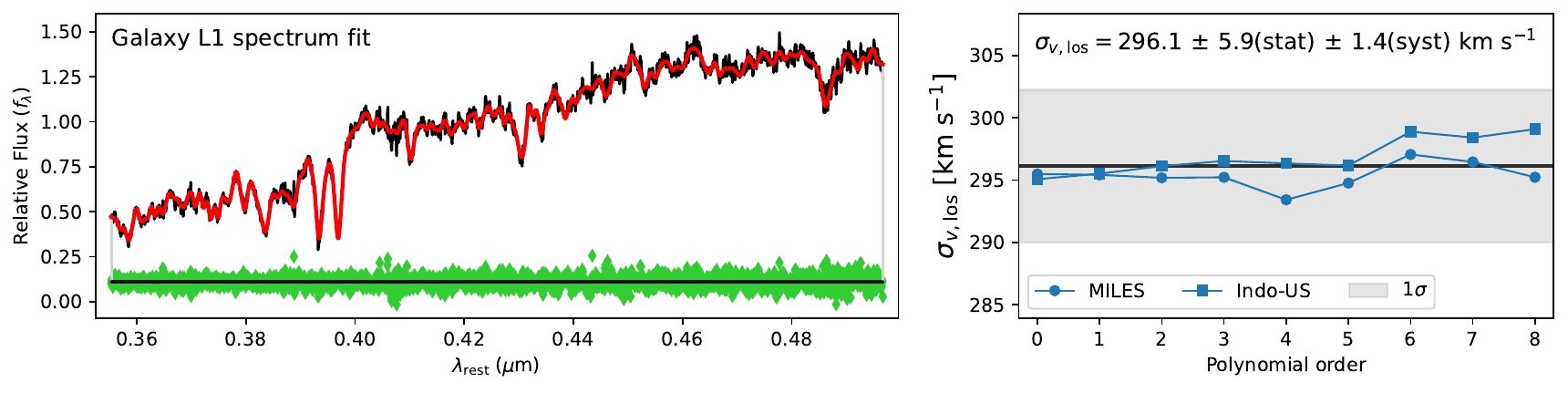}
    \includegraphics[width=\linewidth]{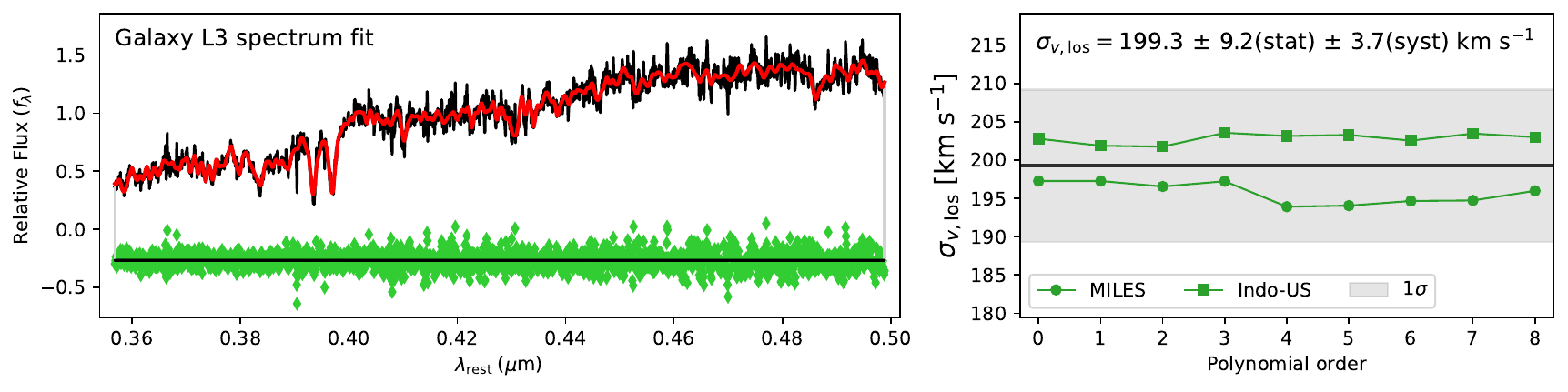}
    \caption{Line-of-sight velocity dispersion \losvd measurements for deflectors L1 (top row) and L3 (bottom row) from the spectra shown in Fig.~\ref{fig:muse_spectra}. \textit{Left panels}: observed spectrum in black, best-fit model in red, and model residuals in green. \textit{Right panels}: best-fit velocity dispersion values for different modeling choices (additive polynomial order and stellar template library). The horizontal black line and the gray shaded area show the mean and $1\sigma$ total uncertainty of $\sigma_{v,\rm los}$, respectively (see Sect.~\ref{sec:kinematics} for details). The statistical (stat) and systematic (syst) uncertainties are also separately indicated at the top of the panel.}
    \label{fig:kinem_L1L3}
\end{figure*}

Fig.~\ref{fig:kinem_L1L3} shows best-fit spectrum models and stellar velocity dispersion measurements for deflectors L1 and L3. We assume \losvd to be normally distributed with mean equal to the average among all best-fit values shown in Fig.~\ref{fig:kinem_L1L3}, which is shown as the horizontal line in the right panels. We then consider two error terms for the variance of this normal distribution. The first term $\sigma^2_{\rm stat}$ is the variance estimated by \ppxf using the diagonal of the parameters covariance matrix at the best-fit position, which corresponds to the statistical error. The second term $\sigma^2_{\rm syst}$ is the variance among of the best-fit values over all polynomial orders and template libraries, and corresponds to the systematic error. The resulting total \losvd uncertainty is simply $\sqrt{ \sigma^2_{\rm stat} + \sigma^2_{\rm syst} }$, which is indicated as a gray shaded area in the right panels of Fig.~\ref{fig:kinem_L1L3}. The individual error terms $\sigma_{\rm stat}$ and $\sigma_{\rm syst}$ are also indicated at the top of these panels. We note that the statistical uncertainty, mainly driven by the S/N of the data, dominates the error budget.

\subsection{Contribution from the cluster \label{ssec:cluster_contrib}}

Since the cluster mass distribution might have a non-negligible effect on the kinematics of L1 and L3, we estimate its impact on $\sigma_{v,\rm los}$ for these two galaxies. Based on the model of \citet{Caminha2023}, we compute the one dimensional projected total mass map without L1, L2 and L3 and measure the cluster mass contribution within the galaxies. Due to the lack of information about the cluster mass profile along the line of sight, we consider two approximations. The first consists of assuming that all cluster mass and galaxies are located at a thin plane. In this case, we use a $0\farcs6$ aperture centered on L1 and L3 (consistently with the measurement aperture) to compute the cluster mass. This estimate is conservative and potentially overestimates the cluster contribution. The second method assumes a mass profile along the line of sight given by a cored isothermal profile \citep[see e.g.,][]{Eliasdottir2007}, that is $\rho(z) \propto 1/(r_{\rm core}^2+z^2)$, where $r_{\rm core}$ is the core radius of the main cluster halo. In this case, the cluster mass is integrated within the same aperture but out to two times the cut radii of L1 and L3 along the $z$ direction. Such assumption of a symmetric profile along the line of sight is still an approximation, but gives a more realistic estimation of the impact on the galaxies kinematics. We obtain that the additional mass of the cluster component increases the galaxies velocity dispersion by factors of 27\% (conservative) or 11\% (realistic) for L1, and only 16\% or 9\% for L3.

\section{Multi-band lens modeling \label{sec:image_modeling}}

\subsection{Data pre-processing and exposure map}

We convert the data flux units from MJy per steradian to electrons per second using the \texttt{PHOTMJSR} keyword (MJy/sr corresponding to one ADU/s) from the data file header and the instrument gain. We retrieve gain values of 2.05 and 1.85 for the short wavelength (SW) and long wavelength (LW) channels, respectively, from the JWST documentation\footnote{\url{https://jwst-docs.stsci.edu/jwst-near-infrared-camera/nircam-instrumentation/nircam-detector-overview/nircam-detector-performance}}). We use the weight map obtained from the drizzling step in order to obtain an effective exposure time per pixel. This allows us to correctly take into account the largely different exposure times throughout the field of view, since \elanz is located right at the edge of certain dithered exposures (see Sect.~\ref{ssec:bayesian_framework}). Since the weight maps provided by the PEARLS team\footnote{\url{https://sites.google.com/view/jwstpearls}} do not contain directly exposure time values, we construct the exposure map as follows. We first retrieve the net exposure time in each filter by dividing the header \texttt{XPOSURE} keyword by the number of detectors of the corresponding NIRCam module (i.e., 8 and 2 for the SW and LW modules, respectively). We then normalize the weight map by dividing it by its maximal value, since we expect the pixel associated to the maximal weight should have the largest exposure time after drizzling. Finally we multiply this normalized weight map by the net exposure time estimated previously, such that the values of the resulting exposure map $t_{\rm exp}$ now ranges from the net exposure time (at maximum) to smaller values for areas with less overlaps between individual exposures. We use this exposure map to estimate the shot noise variance in the data likelihood (see Sect.~\ref{ssec:bayesian_framework}).

\subsection{Point spread function}

\begin{figure}
    \centering
    \includegraphics[width=\linewidth]{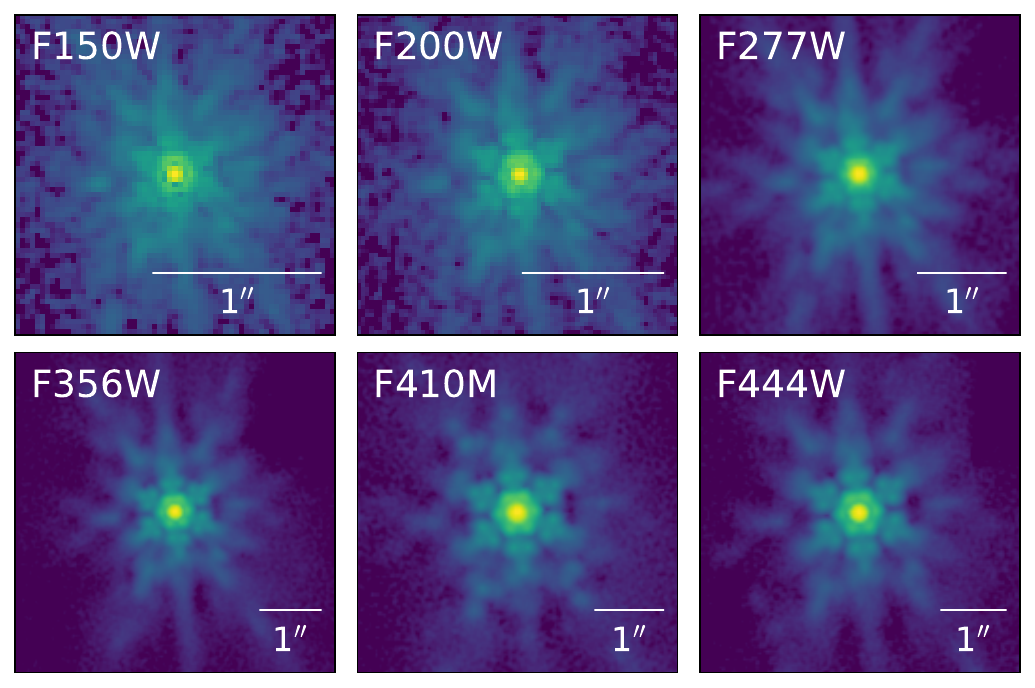}
    \caption{Point spread function (PSF) models for the main NIRCam filters used in this work, constructed from stars detected in the field. Each cutout has been truncated to 30 times the measured resolution (FWHM) in the corresponding filter. The angular size is indiciated by a scale bar at the bottom right of each panel.}
    \label{fig:psf_models}
\end{figure}

Our forward modeling approach incorporates instrumental effects due to the diffraction of light through the telescope by convolving the predicted surface brightness by the point spread function (PSF). We do not rely on a simulated PSF but instead use foreground stars within the field of \elgor to constrain the PSF model, which ensures its consistency with the sampling and orientation of the data. We manually select stars that are not saturated, that are bright enough and that do not significantly overlap with nearby objects. We find 4 stars which fulfill these criteria, located at coordinates (01h02m52.02s, $-$49d14m29.80s), (01h02m49.53s, $-$49d15m52.62s), (01h02m56.19s, $-$49d14m32.95s) and (01h02m54.23s, $-$49d15m02.04s). For band F200W we do not use the last star because it is significantly brighter than the others, such that it can possibly bias the width of the reconstructed PSF. We then extract $201\times201$ cutouts centered on each star in each filter.

We use the public Python software \starred\footnote{\url{https://gitlab.com/cosmograil/starred}} \citep{Michalewicz2023} to construct the PSF model from the selected stars, since \citet{Millon2024} showed that it outperforms other empirical PSF reconstruction methods. We show on Fig.~\ref{fig:psf_models} our PSF model in each NIRCam filter. In \starred, the optimization starts with two dimensional Gaussian profiles that are fitted to each star separately to precisely locate their position within their respective cutout. Then, as an intermediate step, a single Moffat profile, analytically shifted to the positions found in the first step, is jointly fitted to all cutouts simultaneously. The final step consists in optimizing a pixelated grid to fit all PSF features that cannot be captured with analytical profiles. For this step, the Moffat component is held fixed to the best-fit found in the second step. The pixelated background component is regularized with sparsity constraints in wavelet domain. We refer the reader to \citet{Michalewicz2023} and the online documentation for more details about the algorithm. Overall, the resulting PSF model is composed of a Moffat profile superimposed with pixelated deviations that capture local features such as Airy rings and diffraction spikes. \starred being based on the automatic differentiation library JAX \citep{jax2018github}, it can optimize the large number of parameters (of the order of $201^2\sim10^4$) using efficient gradient-based algorithms. It takes approximately 3 minutes on a personal laptop to obtain a PSF model in one filter.

\subsection{Differentiable multi-band lens modeling \label{ssec:lens_modeling}}

We simultaneously model several JWST/NIRCam bands in which both the main arc and its counter image are visible. As each band in the reduced PEARLS data set is already aligned, we do not add any additional degree of freedom to correct for possible misalignment between the bands. In this work, we explicitly model the mass distribution of L1, L2, and L3 galaxies as main deflectors and take into account the non-uniform mass distribution of the \elgor cluster. The origin $(0'',0'')$ of our model coordinate system coincides with WCS coordinates $(15.7057851^\circ, -49.2518014^\circ)$ in the JWST/NIRCam data from PEARLS \citep{Windhorst2023_PEARLS}. The pixel size in all bands is $0\farcs03$, the $x$ axis is positive towards the West, and the $y$ axis is positive towards the North.

As the redshifts of L1 and L3 are slightly different (0.872 and 0.864, respectively), we check if these deflectors should be placed on different lens planes. We use the multi-lens plane interface of \lenstro \citep{Birrer2018lenstro,Birrer2021lenstro} to estimate the error on the deflection field when assuming a single lens plane at the redshift of L1. More specifically, we compute the maximal deflection angle difference between single-plane and multi-plane deflection fields over the arc region. We find that the deflection angle changes only by $0\farcs0098$, which is $6.3$ times smaller than the JWST resolution in band F150W \citep[${\rm FWHM_{\rm F150W}}=0\farcs062$ from Table 1 of][]{Windhorst2023_PEARLS}. Therefore, we do not employ the multi-lens plane formalism in this work, and essentially assume that L1, L2 and L3 are in the same lens plane.

Our baseline model of the deflectors mass distribution is composed of an elliptical power-law (EPL) profile \citep{Tessore2015} for L1 and L2, and singular isothermal ellipsoid (SIE) profile for L3. Each of these profiles are parametrized by an Einstein radius \thetaE, logarithmic (3D) mass density slope $\gamma$ (an SIE profile has $\gamma=2$), axis ratio $q_{\rm m}$, position angle $\phi_{\rm m}$ and centroid $(x_{\rm m}, y_{\rm m})$. In addition, we include in some of our models an external shear field to capture additional angular structures in the mass distribution. The external shear is parametrized by its strength $\psi_{\rm ext}$ and orientation $\phi_{\rm ext}$, with origin centered on our coordinate system.

We model the surface brightness of the deflectors using a combination of Sérsic profiles \citep{Sersic1963}. The Sérsic profile is parametrized by its half-light radius $\theta_{\rm eff}$, Sérsic index $n_{\rm s}$, axis ratio $q_{\rm \ell}$, position angle $\phi_{\rm \ell}$, centroid $(x_{\rm \ell}, y_{\rm \ell})$ and amplitude $I_{\rm eff}$ at $\theta_{\rm eff}$. We additionally include a uniform light profile to the lens light model in order to take into account potentially over- or under-subtracted sky level, parametrized by independent intensities $I_{\rm bkg}$ in each band.

For the source surface brightness we use a multi-band pixelated model based on a Gaussian process defined in Fourier space with non-parametric covariance. We refer to \cite{Arras2022} for a complete description of the model, and briefly describe it here for completeness. Our forward source model $\boldsymbol{s}$ can be written as
\begin{align}
\label{eq:src_correl_field}
    \boldsymbol{s} = \exp\bigg[ \boldsymbol{F}^{-1} \big( \boldsymbol{A} \odot \boldsymbol{\xi} \big) + \boldsymbol{\delta} \bigg] \ ,
\end{align}
where $\boldsymbol{\xi}$ is a so-called excitation field, $\boldsymbol{A}$ is an amplitude operator, which is the square root of the power spectrum, $\boldsymbol{F}^{-1}$ is the inverse Fourier transform, $\boldsymbol{\delta}$ is a uniform offset (in real space) and $\odot$ is the point-wise multiplication. To enforce positivity of the source pixel values, we take the exponential of the Gaussian process. The number of excitation field parameters is $n_\lambda \times n_{\rm src}^2$, where $n_\lambda$ is the number of bands and $n_{\rm src}$ is the number of source pixels along each spatial dimensions.

%\footnote{We note that the model defined in Eq.~\ref{eq:src_correl_field} assumes periodic boundary conditions along both spatial and wavelength dimensions. In source plane---whose extent is dynamically adapted using the arc mask---, the source lies well within the spatial extent of the field such that periodic conditions do not impact the morphology of the source. However, along the wavelength dimension, we effectively increase $n_\lambda$ by 4 extra dimensions which are simply dropped when evaluating the source model, in order to prevent introducing unphysical correlations between the bluer and redder bands.}

We use the lens modeling code \herculens\footnote{\url{https://github.com/Herculens/herculens}} \citep{Galan2022} that implements the various model components, computes lensing quantities, performs ray-tracing and convolutions with the PSF. For the correlated field model, we use the software package \nifty\footnote{\url{https://gitlab.mpcdf.mpg.de/ift/nifty}} \citep{nifty1,nifty3,nifty5}. More specifically, we use the \jax implementation \textsc{NIFTy.re} \citep{niftyre}, since \herculens is also based on \jax. Our full modeling and inference pipeline is thus differentiable, can be pre-compiled and runs on both CPUs and GPUs.

\subsection{Constraints from stellar kinematics \label{ssec:kinem_to_mass}}

\begin{figure}
    \centering
    \includegraphics[width=0.9\linewidth]{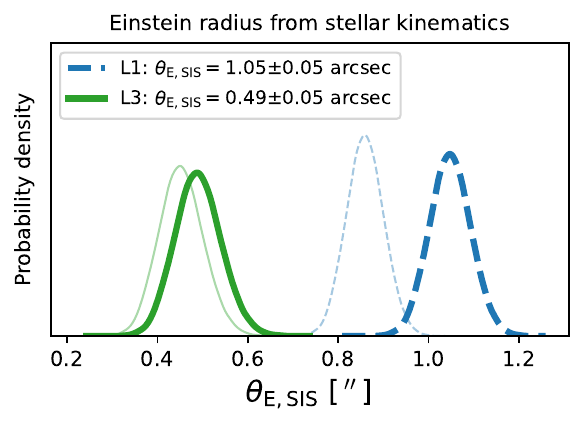}
    \caption{Probability distributions for the Einstein radius \thetaE of deflectors L1 and L3, based on \losvd measurements shown in Fig.~\ref{fig:kinem_L1L3}, assuming an isothermal (SIS) mass distribution and taking into account the contribution from the host cluster. For comparison, the faint distributions are obtained with the more conservative contribution from the cluster (see Sect.~\ref{ssec:cluster_contrib}). For each samples from the \losvd probability distribution, we compute \thetaE following Eq.~\ref{eq:veldisp2thetaE}. In this work we only use the Einstein radius estimate for L3 to complement lensing constraints, and only show the estimate for L1 for completeness.}
    \label{fig:theta_E_priors}
\end{figure}

\begin{table}
    \caption{Dynamical properties of cluster members L1 and L3.}
    \label{tab:props_from_muse}
    \renewcommand{\arraystretch}{1.4}
    \centering
    \small
    \begin{tabular}{lcc}
        \hline\hline
        Galaxy & L1 & L3 \\
        \hline
        $z$      & $0.8722 \pm 0.0001$ & $0.8642 \pm 0.0002$ \\
        $\losvd$ [\ks] & $296.1 \pm  6.1$    & $199.2 \pm 10.0$    \\
        $\thetaESIS$ [\,$''$\,], realistic\tablefootmark{$\dag$}    & $1.05 \pm 0.05$ \tablefootmark{$\star$} & $0.49 \pm 0.05$ \\
        $\thetaESIS$ [\,$''$\,], conservative\tablefootmark{$\dag$} & $0.86 \pm 0.04$ \tablefootmark{$\star$} & $0.45 \pm 0.05$  \\
        \hline
    \end{tabular}
    \tablefoot{All quantities are inferred from VLT/MUSE data only, and correspond to the redshift $z$, LOS stellar velocity dispersion $\losvd$ and corresponding Einstein radius $\thetaESIS$ assuming an SIS mass distribution (Eq.~\ref{eq:veldisp2thetaE}). Uncertainties for redshifts and velocity dispersions contain both statistical and systematic uncertainties, adding in quadrature.\\
    \tablefoottext{$\dag$}{The ``realistic'' and ``conservative'' cases relate to the estimated contribution of the cluster, described in Sect.~\ref{ssec:cluster_contrib}.}\\
    \tablefoottext{$\star$}{In this work, we infer that the density profile L1 is steeper than for an SIS from lensing constraints (Sect.~\ref{ssec:mass_constraints}).}
    }
\end{table}

We estimate the Einstein radius of L1 and L3 from their LOS velocity dispersion by assuming a mass distribution following a singular isothermal sphere (SIS) and using the relation
\begin{align}
    \label{eq:veldisp2thetaE}
    \thetaESIS = f_{\rm cluster} \times 4\pi\frac{\losvd^2}{c^2}\frac{\Dds}{\Ds} \ ,
\end{align}
where $c$ is the speed of light, \Ds and \Dds are angular diameter distances to the source and between the deflector and the source, respectively. We add a multiplicative factor $f_{\rm cluster}$ to take into account the cluster contribution to the gravitational potential. As discussed in Sect.~\ref{ssec:cluster_contrib}, the cluster realistically contributes to the mass of L1 and L3 at the $11\%$ and $9\%$ levels, respectively, hence we set $f_{\rm cluster,L1} = 1 - 0.11$ and $f_{\rm cluster,L3} = 1 - 0.09$. For angular diameter distances, we compute \Ds and \Dds corresponding to L1 and L3 using redshift estimates from \ppxf fits (i.e., $z_{\rm L1}=0.872$ and $z_{\rm L3}=0.864$), and assume $\zs=2.291$ for the source redshift. The resulting probability distributions lead to $\theta_{\rm E,L1}=1\farcs05\pm0\farcs05$ and $\theta_{\rm E,L3}=0\farcs49\pm0\farcs05$, as shown in Fig.~\ref{fig:theta_E_priors} and compared to the distributions for a more conservative cluster contribution. We summarize in Table~\ref{tab:props_from_muse} the redshift, velocity dispersion and Einstein radius estimates.

We emphasize that, to complement lensing constraints, we only use the Einstein radius estimate for L3 consistently with the assumption of an isothermal density profile used in the lens model. As shown in \citet{Kamieneski2023} and in Sect.~\ref{ssec:lens_modeling}, the lensing data is not constraining enough to probe the mass distribution of L3 beyond the assumption of an isothermal profile. For L1 however, we find~\ref{ssec:lens_modeling} that the mass density slope at the Einstein radius is steeper than isothermal.

\subsection{Bayesian framework \label{ssec:bayesian_framework}}

\begin{figure}
    \centering
    \includegraphics[width=\linewidth]{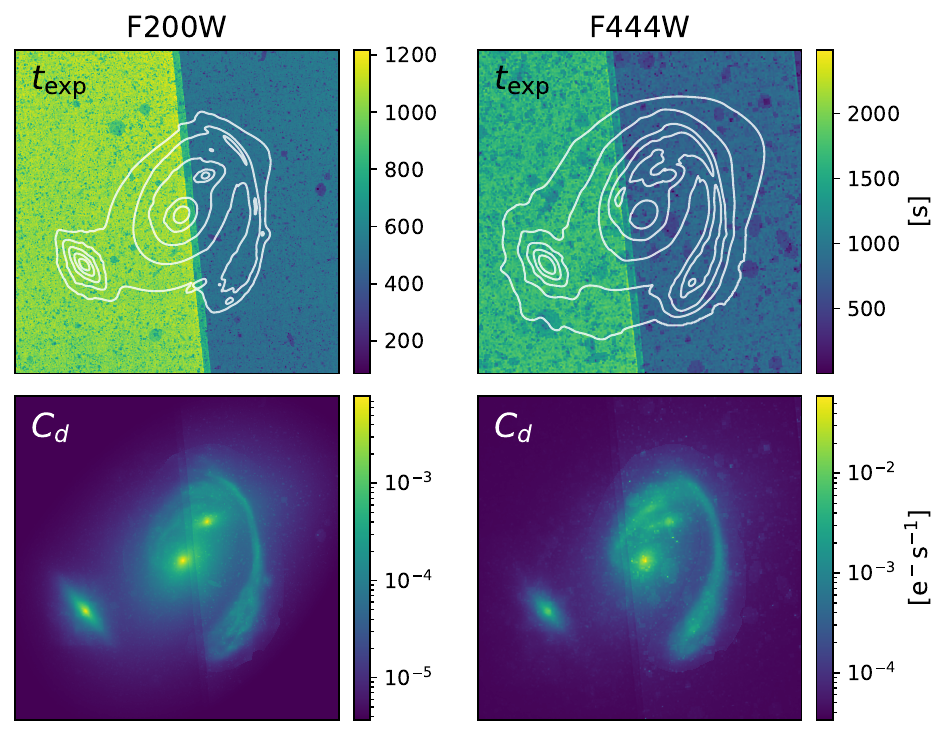}
    \caption{Non-uniform exposure time over the field of view of \elanz due to the combination of dithered exposures and masking of cosmic rays. \textit{Top row}: example exposure maps for filters F200W and F444W. White isophotes indicate the position of the arcs and deflectors with respect to features in the exposure map. \textit{Bottom row}: diagonal of the data covariance matrix (i.e., the noise map) estimated from the exposure map and the model-predicted flux to compute the shot noise contribution. The detector edge and cosmic ray imprints are still clearly visible in the noise map and are taken into account during modeling.}
    \label{fig:exp_coverage}
\end{figure}

We cast the optimization and inference problem in a Bayesian framework to combine the imaging and kinematics datasets. We denote the imaging likelihood $\mathcal{P}\big(\boldsymbol{d}_{\rm img}\,|\,\boldsymbol{m}_{\rm img}(\Theta)\big)$, where $\boldsymbol{d}_{\rm img}$ is the imaging data and $\boldsymbol{m}_{\rm img}$ is the model with parameters $\Theta$. We further assume that data noise follows a normal distribution with diagonal covariance matrix $C_{\rm d}$, which results in the following analytical formula for the log-likelihood:
\begin{align}
    \label{eq:data_likelihood}
    \nonumber
    \log\,\mathcal{L}\big(\Theta\big) &\equiv \log\,\mathcal{P}\big(\boldsymbol{d}_{\rm img}\,|\,\boldsymbol{m}_{\rm img}(\Theta)\big) \\
    \nonumber
    &= \ - \frac12\,\bigg[\boldsymbol{d}_{\rm img} - \boldsymbol{m}_{\rm img}(\Theta)\bigg]^\top C_{\rm d}^{-1}(\Theta)\, \bigg[\boldsymbol{d}_{\rm img} - \boldsymbol{m}_{\rm img}(\Theta)\bigg] \\
    &\ \ \ \ \ + \log\bigg( 2\pi\sqrt{\det C_{\rm d}(\Theta)}\bigg) \ .
\end{align}
The first term in Eq.~\ref{eq:data_likelihood} is simply $-\chi^2/2$. The diagonal of the covariance matrix $C_{\rm d}$ represents the noise variance per data pixel, which is the combination of a constant background noise and the shot noise due to the flux of the object itself. When evaluating Eq.~\ref{eq:data_likelihood}, we estimate the shot noise from the model-predicted flux, hence the dependence of $C_{\rm d}$ on the model parameters $\Theta$. In the case of JWST data of \elanz, the data covariance is non trivial because the lens is located at the edge of the detector on some of the individual exposures, as visualized in Fig.~\ref{fig:exp_coverage} on which a diagonal discontinuity is clearly visible after drizzling. We take into account this spatially varying exposure time over the field of view by self-consistently updating $C_{\rm d}$ during model optimization. Since the data has units of electrons per second, the shot noise variance in $C_{\rm d}$ corresponding to pixel $i$ is $m_{\rm img,i}/t_{\rm exp,i}$.

We incorporate in the inference the constraints from stellar kinematics data $\boldsymbol{d}_{\rm kin}$ measured in Sect.~\ref{ssec:kinem_to_mass} by imposing a Gaussian prior on the Einstein radius of L3:
\begin{align}
    \mathcal{P}(\theta_{\rm E, L3}) \equiv \mathcal{P}(\theta_{\rm E, L3}\,|\,\boldsymbol{d}_{\rm kin}) \sim \mathcal{N}\left(0\farcs49, 0\farcs05\right) \ ,
\end{align}
which corresponds to the distribution shown in Fig.~\ref{fig:theta_E_priors}.

The probability distribution of the source excitation field follows the standard multi-variate normal distribution. Both source power-spectra (along the spatial and wavelength dimensions) are power-laws in logarithmic scale parametrized by a log-normal prior on the fluctuations amplitude, and a Gaussian prior on the power-law slope. In addition, a global additive offset is added, parametrized by a mean and standard deviation, the latter itself following a log-normal distribution with free mean and width parameters. For the formal Bayesian formulation of the correlated field model, we refer the reader to \citet{Arras2022}. Other lens mass parameters are sampled according to either normal or uniform distributions, which we detail in Sect.~\ref{ssec:modeling_seq}.

From a set of different model variations within a given model family, we further follow Bayesian principles and use the Bayesian information criterion (BIC) to weight the different models before marginalization. The BIC is defined as
\begin{align}
    \label{eq:def_bic}
    {\rm BIC} = n_{\rm p}\log(n_{\rm d}) - 2\log \mathcal{L}(\hat{\Theta}) \ ,
\end{align}
where $n_{\rm p}$ is the number of optimized model parameters, $n_{\rm d}$ is the number of data points used to constrain the model and $\hat{\Theta}$ is the set of model parameters that maximize the likelihood function $\mathcal{L}$ defined by Eq.~\ref{eq:data_likelihood}. After computing the BIC values associated to each model variation, we can compute the associated (relative) weights based on their BIC difference with the model having the lowest BIC value (i.e., the highest Bayesian evidence). We use the BIC as a proxy for the computationally expensive Bayesian evidence; however, the evidence lower bound (ELBO) used in variational inference can also be used in principle.

At this point, it is important to note that realistically, one can only sparsely sample the space of all possible model variations within a model family. Therefore, we must take into account this sparse sampling and correct the weights by using the scatter in BIC values estimated over the specific set of model variations. We correct the weights using a strategy inspired by recent strong lensing analyses of lensed quasars \citep[e.g.,][]{Rusu2020,Shajib2020}. In particular, we convolve the relative weights based purely on BIC differences by a Gaussian window function whose width is based on the standard deviation of the BIC values \citep[e.g., see Eq. 12 from][]{Rusu2020}. Not correcting for sparse sampling would typically result in only one or two individual models to contribute to the final posterior, which could lead to severe underestimation of parameters uncertainties.

\subsection{Modeling sequence \label{ssec:modeling_seq}}

From the original 8 filters of the \elgor data set obtained by PEARLS, we select 6 in which the lensed arcs are clearly visible, namely we discard filters F090W and F115W. These filters also have overall lower S/N with significant non-Gaussian noise patterns. Therefore, we use data in filters F150W, F200W, F277W, F356W, F410M and F444W to constrain our lens models.

We start by modeling the surface brightness of the three deflectors L1, L2 and L3. Based on the NIRCam imaging data these are likely elliptical galaxies, although L2 and L3 have high ellipticities which could be the hint of a disk component. As all three deflectors are very bright and display significant deviations from axisymmetry, several model components are required to model their light distribution to acceptable levels. In particular we use 2 concentric Sérsic profiles for L1 and L2, and 3 concentric Sérsic profiles for L3. We also include a free constant background light component to properly account for any systematic offset from previous background subtraction steps. We do not model possible intra-cluster light (ICL), which is clearly visible only closer to the main cluster components in the JWST data. \elanz is located at a projected distance of $\sim 230$ kpc from the brightest cluster member (BCG), in a region where the ICL is fainter. Moreover, as estimated by \citet{Diego2023}, at a projected distance of 100 kpc from the BCG, the ICL in \elgor contributes to $\lesssim1\%$ of the projected mass. If a significant fraction of ICL was missing in our model, it would appear in the model residuals as a large scale gradient over the field of view, consistently between the filters. As we do not observe such large scale residuals (see also Appendix~\ref{app:ssec:lens_light}), we do not further increase the flexibility of our light model. To maximize the fit quality in each of the six JWST filters and given the limited flexibility of the Sérsic profiles, we optimize lens light parameters in each bands separately. This single-band fitting step also allows us to measure the scatter in the position of the deflectors light distribution, and further use it as a prior for the centroids of their mass distribution. For each filter, we carefully design a mask to exclude the region containing significant source flux from the likelihood. We also exclude several other luminous objects in the field of view, mostly small galaxies or potential tidal stripping features. The optimization of all lens light parameters is performed using the second-order gradient descent algorithm BFGS \citep{Nocedal2006}, accessible in \herculens as a wrapper for the python package \textsc{JaxOpt} \citep{jaxopt_implicit_diff}. All lens light parameters are then fixed to their best-fit values for subsequent modeling steps.

We first search for an approximate mass model to use as a suitable starting point later on for the full multi-band modeling. The lensing features of \elanz being complex, we could not obtain such a mass model by fitting only simple light components such as Sérsic profiles in source plane. Instead, we construct a slightly more elaborate source model by combining a Sérsic profile with a shapelets basis set \citep{Birrer2017}. The shapelets model is implemented in \herculens \citep{Galan2022} as a wrapper around the \jax implementation of \textsc{GigaLens} \citep{Gu2022}, itself using \lenstro routines \citep{Birrer2018lenstro,Birrer2021lenstro}. We fix the lens light model and optimize all remaining parameters using the BFGS algorithm.

We then replace this intermediate source model with the 3D correlated field described in Eq.~\ref{eq:src_correl_field}. The arc mask is updated such that it encloses all the flux from the source galaxy, such that it covers areas that contain no source flux as well (ensuring that the full extent of the source is being reconstructed). Given a set of mass model parameters, this arc mask dynamically sets the extent of the pixelated grid in the source plane, such that the source grid pixel scale relative to the angular scale of the source remains approximately constant\footnote{A similar strategy is used in the lens modeling software GLEE \citep{SuyuHalkola2010, Suyu2012}.}. Beside the arc mask, the overall likelihood mask remains the same as for the lens light fitting step. We converge on a fiducial number of source pixels $n_{\rm src}=100$ along each spatial dimension, by progressively increasing $n_{\rm src}$ until we observe no significant changes in the residuals (i.e., the reduced $\chi^2$ remains approximately constant). We also find that increasing $n_{\rm src}$ does not significantly affect the morphology of the lensed galaxy, but only allows to capture more of the compact clumps within its disk, which only cover a negligible fraction of pixels entering the $\chi^2$ calculation. We maintain the lens light parameters fixed to their best-fit values, and jointly fit for the source and lens mass parameters. To allow our model to compensate for imperfect lens light modeling, we add a free uniform amplitude over the arc mask region, independently in each band.

For the mass distribution of L3, we fix all its SIE parameters to their corresponding lens light parameters, except for the Einstein radius as it is the only parameters that can be realistically constrained given its distance to the arcs. For other mass model parameters---center, position angle and axis ratio---we impose broad priors informed by the best-fit values of their lens light model counterpart, although not unreasonably broad to keep the randomly drawn initial parameter values within a physically plausible range. The position angles of the two EPL profiles are sampled from a Gaussian prior with mean centered on the best-fit value from the reference filter (F444W with higher S/N) and a width of 10 degrees. The axis ratio $q_{\rm m}$ of the EPL profiles are sampled from a uniform prior in the range $[{\rm max}(0.5,\, q_{\ell, \rm ref} - 3\sigma_q), 1]$, where $q_{\ell, \rm ref}$ is the axis ratio in the reference filter and $\sigma_q$ is the scatter of axis ratio over the filters. This sensible prior on $q_{\rm m}$ is motivated by previous results that the dark matter, thus the total, mass ellipticity of elliptical galaxies is correlated with their light ellipticity \citep[e.g.,][although such correlations may be impacted by selection functon effects]{Dubinski1994,Sluse2012,Shajib2021}. In our case however, our generic choice of axis ratio prior effectively leads to a uniform prior over the interval $[0.5,\, 1]$ for both L1 and L2. We also allow the center of EPL profiles to vary and assign Gaussian priors with mean and width equal to mean and standard deviation of the positions across the filters of the corresponding light distribution. Thus, all our models allow for a misalignment between the light and total mass distribution of the main deflectors. The logarithmic slopes of the EPL profiles have wide Gaussian priors centered on $\gamma=2$ with $10\%$ standard deviation, wide enough not to impact the posterior distributions. Finally, for models that include an external shear field, we draw elliptical shear components ($\gamma_{\rm ext, 1}$, $\gamma_{\rm ext, 2}$) from a Gaussian centered on zero with width 0.05. We check that our inferred results are not prior-dominated by verifying that the aforementioned priors are wider or comparable to the tighter posteriors constrained by the multi-band JWST data.

We use variational inference (VI) to model the joint posterior distribution over the numerous model parameters. More specifically, we use geometric variational inference \citep[\textsc{geoVI},][]{Frank2021_geovi}, which is able to capture non-linear covariances even in large dimensions. Due to restricted GPU memory, we maintain the number of individual samples used to estimate the Kullback-Leibler divergence (which is the metric optimized in VI) to a relatively low value (we use 32 samples for the last five \textsc{geoVI} iterations). However, all samples are fully independent and are drawn using antithetic sampling, which leads to a larger effective number of samples and a reduced sampling variance \citep[see Eq.~65 from][]{Knollmuller2019_mgvi}. We take into account possible systematic errors by performing different optimization runs within a given model family, which we refer to as model variations. We obtain a well-sampled final posterior distribution by approximating the marginalized posterior with a multivariate normal distribution, from which we draw an arbitrary number of samples. The mean and covariance matrix of this normal distribution is computed from the samples of individual model variations, weighted by the relative BIC differences computed as described in Sect.~\ref{ssec:bayesian_framework}. We emphasize that while the final joint posterior distribution is, by construction, a linear approximation of the underlying distribution, it is built from \textsc{geoVI} samples that individually capture the non-linear covariances in the parameter space \citep{Frank2021_geovi}.

The lens models considered in this work have $\sim10^5$ parameters\footnote{The ``effective'' number of model parameters, in particular due to the field regularization of the source model, is much lower (although it is not explicit).} and are constrained by $\sim 10^6$ pixels (excluding pixels outside the masked areas). We note that just-in-time compilation, gradient-based inference algorithms and GPU parallelization all lead to a low computation time. In particular, it takes approximately 140 minutes on a single NVIDIA A100 GPU to obtain one variational inference estimate of the joint posterior distribution, for a given model variation. Increasing the resolution of the source, even significantly, only marginally impacts runtime.

\subsection{Model families \label{ssec:model_families}}

We model the JWST data set assuming three different model families---which can be seen as two intermediary models and one final model---depending on the description of the nearby cluster \elgor. We first consider a ``Shear only'' model, based on the assumption that \elgor's deflection field can be approximated by a single external shear component over a field of view of $\sim10''$ centered on L1. This assumption follows the recent work of \citet{Kamieneski2023}, and is also the most common assumption to model nearby massive structures external to the main deflectors.

Instead of a uniform external shear, we build a ``Cluster only'' model by using one of the most recent lens models of \elgor. We use the model of \citet{Caminha2023} based on HST imaging data and MUSE spectroscopic data (the same data we use in Sect.~\ref{sec:kinematics}) that allowed the authors to spectroscopically confirm the 23 multiply imaged sources used to constrain the model. This cluster model is parametrized by a set of dual pseudo-isothermal mass density (dPIE) profiles associated to the 243 cluster members and 20 foreground group members, as well as pseudo-isothermal elliptical mass density (PIEMD) profiles associated to each of the two dark matter cluster-scale components of the cluster. For the complete description of the mass model, see \citet{Caminha2023}. We note that two other cluster-scale lens models have been recently published by \citet{Diego2023} and \citet{Frye2023}, also using the JWST imaging data as in our work. The spectroscopic model of \citet{Diego2023} is based on the same the lensing constraints as \citet{Caminha2023} but follows a non-parametric modeling approach. Such a non-parametric model is expected to be more accurate close to multiple image pairs, but may be less robust further away from these constraints, which is the case at the location of \elanz \citep[no constraints from \elanz were used in the spectroscopic model of][]{Diego2023}. The model of \citet{Frye2023} is based on the LTM approach, new photo-$z$ measurements and incorporates constraints within the arc of \elanz. Since we do not use the mass distribution of L1, L2 and L3 from the cluster-scale model, we expect marginal differences between \citet{Caminha2023} and \citet{Frye2023} models over the field of view of \elanz. Therefore, in this work we consider the model of \citet{Caminha2023} after removing the mass components associated to the three deflectors.

The third ``Cluster+Shear'' model is a combination of the two models above as it contains both an external shear component and the cluster's deflection field. We consider this more advanced model as our fiducial one. The external shear in this case is expected to capture additional azimuthal structures that have not been included in the cluster lens model, such as nearby faint galaxies that are visible in some NIRCam filters. However, as recently recalled by \citet{Etherington2023_extshear}, there is no guarantee that external shear captures only structures external to the lens, but can instead compensate for the lack of azimuthal freedom in the main deflectors mass model. In our case, one can also interpret the external shear as a local correction to the cluster deflection field.

\section{Results \label{sec:results}}

\subsection{Overall fit to the multi-band NIRCam data}

\begin{figure*}
    \centering
    \includegraphics[width=0.85\linewidth]{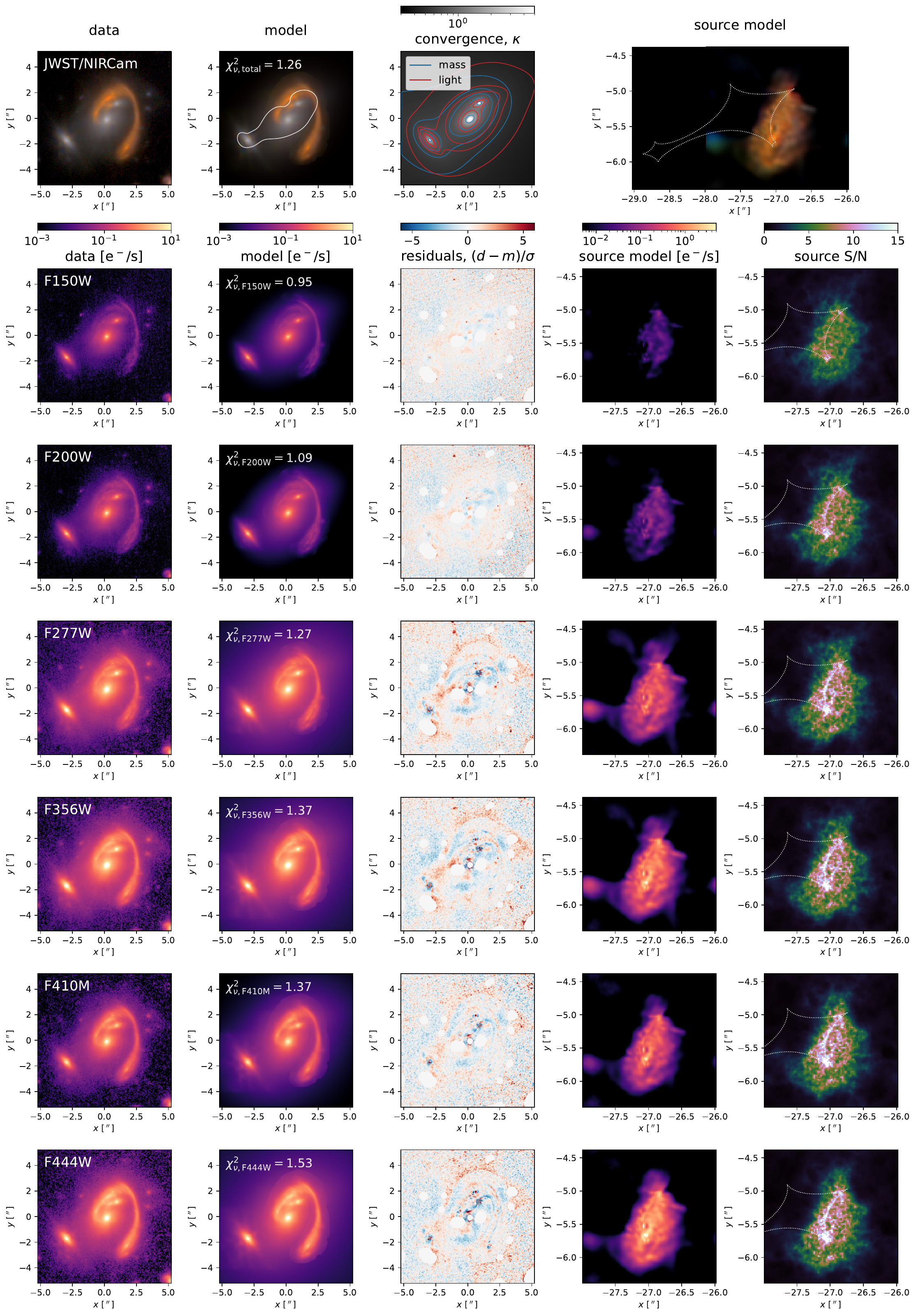}
    \caption{Mean lens model among the best-fit posterior samples for our fiducial ``Cluster+Shear'' model. \textit{First row, from left to right}: color image of the data, color image of the model with reduced $\chi^2$ from image residuals (which differs from the full log-likelihood, see Eq.~\ref{eq:data_likelihood}), convergence map (with convergence and surface brightness contours), and color image of the unlensed source model. \textit{Remaining rows, from left to right}: data in the indicated filter, model, normalized residuals (areas excluded from the fit appear white), unlensed source model, S/N source map defined as the mean source model divided by the standard deviation of the model posterior in each source pixel. Critical lines and caustics are indicated in some of the panels as white solid line in the image model panel, and dotted lines in the right column panels. The source S/N is higher within and along the caustics, since image multiplicity and lensing magnification are higher, respectively.}
    \label{fig:bestfit_lens_model_summary}
\end{figure*}

Among the model families we consider, the ``Cluster+Shear'' model provides the best fit to the multi-band imaging data, with a reduced chi-square of $\chi^2_\nu=1.25$ computed among all fitted data pixels. The ``Shear only'' model leads to a slightly worse fit with a $\chi^2_\nu$ increase of $\gtrsim0.01$, followed by the ``Cluster only'' with a further increase of $>0.02$. We show in Fig.~\ref{fig:bestfit_lens_model_summary} an instance of the best ``Cluster+Shear'' model of our baseline setup (i.e., a source model with $100$ pixels along each spatial dimensions). As we optimize model parameters using VI, Fig.~\ref{fig:bestfit_lens_model_summary} only shows the model corresponding to the mean among the best-fit VI posterior samples (see caption for the description of each panel). Moreover, we can also use these same samples to divide the mean by the standard deviation of the posterior, which results in a S/N map that we show for the source plane in the rightmost column of Fig.~\ref{fig:bestfit_lens_model_summary}.

In the image plane, the model fits remarkably well the data in all bands. The lensed arcs, although containing a large number of features at difference scales, are particularly well fitted. We also note the dynamic range of the source model, which spans more than three orders of magnitudes from fainter structures in filter F150W to the brighter central region in filter F444W. The surface brightness of the L2 and L3 galaxies are also accurately modeled. Small scale residuals that still remain close to L2 and L3 centroids are likely caused by slight inaccuracies in the PSF model, as they are similar to those of quasar-host separation analyses (e.g., see Fig.~2 of \citet{Ding2022} or Fig.~4 of \citet{Stone2023}).

We distinguish two main categories of residuals. First, the surface brightness of L1 is not accurately reproduced by the model close to the center of the galaxy and beyond the arc in the norther part of the cutout (see Appendix~\ref{app:ssec:lens_light}). Second, the brightest part of the southern arc (corresponding to the very center of the source galaxy), as well as the compact feature (almost point-like) at the north of L2, are not fully captured by the source model. While the southern part leads to mild residuals only in F444W (below the $3\sigma$ level), the compact feature leads to strong residuals (more than $6\sigma$ within some pixels) with some symmetric features. In the source plane, this clump is clearly offset and distinct from the center of the source galaxy, as shown in the rightmost panel of Fig.~\ref{fig:model_color}. We believe this clump is resolved (as opposed to a point source), because increasing the source model resolution allows us to capture its extent over several pixels. We discuss in more detail this specific feature in Sect.~\ref{ssec:clump_detection}.

The multi-band model of the background galaxy reveals a heavily wavelength-dependent morphology. In F150W and F200W filters, most of the detected regions of the source are located outside the main astroid (a.k.a., tangential) caustic. Interestingly, the morphology in these filters do not reveal any significantly bright component that could be associated to a centroid. It is only in F277W and redder bands that we see a more symmetric morphology containing features that could be attributed to a brighter central bulge, dust attenuated areas, or regions with lower stellar density. Moreover, our model reveals new features located south of the bulge, which may be associated to spiral arms.

The source S/N is, as expected, higher along the caustics, because these regions are highly magnified and have at least two multiple images visible in the data. The bulge of the source galaxy is the highest S/N region and is located remarkably close to the southern part of the main tangential caustic (which features a double-cusp shape). Coincidentally, the upper part of the source crosses the northern cusp of the same caustic. Overall, the source S/N is larger inside and along the caustics, but is not significantly lower outside (as one may expect). We attribute this globally uniform S/N to the significant brightness outside the caustics in the mid-infrared, which effectively leads to higher S/N at these locations. We also investigate if some features in the source S/N map can be a consequence of the spatially varying exposure time (see Fig.~\ref{fig:exp_coverage}) and find that lower S/N regions (after being lensed) do not overlap with lower exposure time regions. Therefore, the source S/N is dominated by lensing magnification and the intrinsic source brightness, and the non-uniform exposure time does not impact the reconstructed source.

Our source S/N maps also allow us to discard highly uncertain source plane features, such as the smooth blob located around $(-27\farcs8,-5\farcs6)$. These spurious features are in reality caused by the leakage of un-modeled lens light features that contaminate the source model. Indeed, the purely smooth light model of L1 is inaccurate in some locations that overlap with the arc mask, causing the lensed source model to predict non-zero flux along the edge source plane grid. However, the posterior uncertainty of the source pixels is large, as expected. We expect a better model of the deflectors surface brightness to remove these low-significance features from the source model. While a more flexible deflectors light model may be warranted in the future, our current model does not support the hypothesis of possible over- or under-subtraction of deflectors light over the arc mask region (i.e., all additional amplitudes varied over the arc mask region in each band remain extremely low $\lesssim10^5$ e$^-$/s).

While designing the arc mask, we noticed one luminous structure on the West side of the arc around image plane position $(2\farcs8, -0\farcs7)$ appearing bluer than the rest of the arc in the NIRCam data. This color difference may be the sign that this structure does not belong to the source, but is located along the line-of-sight instead. We also notice that this bluer structure has similar colors as the two galaxies located West to the arc (visible in the color image of Fig.~\ref{fig:model_color}). Moreover, the foreground ($z=0.63$) galaxy group reported in \citet{Caminha2023} contains galaxies that appear bluer than \elgor members. These blue objects in the vicinity of \elanz may thus be low-mass or dwarf galaxies lying at a similar redshift as the foreground galaxy group. Unfortunately, these objects are not detected in the MUSE data, such that we are unable to confirm their nature. Therefore, due to its close proximity with the arc and the absence of objective evidence that it does not belong to the source, we did not mask it and thus allow our source model to capture it. The unlensed version of this structure is clearly visible in the fourth column of Fig.~\ref{fig:bestfit_lens_model_summary} as a thin elongated feature of size $\sim0\farcs1$ around position $(-26\farcs4, -5\farcs2)$. We note that \citet{Kamieneski2023} also did not mask this structure and ray-traced it to their source plane model (visible on the top right panel of their Fig. 2). Moreover, this structure being faint in the data, it has likely a small mass which may explain why we do not see signs of local distortions in the arc.

\subsection{Critical lines and caustics}

\begin{figure*}
    \centering
    \includegraphics[width=\linewidth]{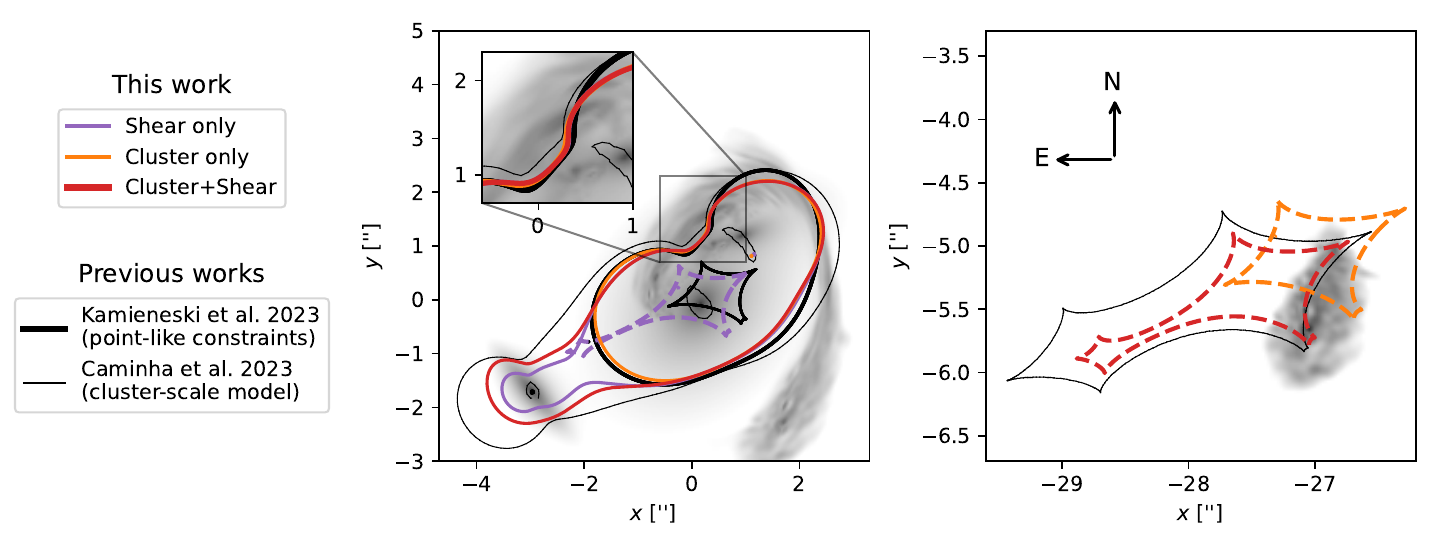}
    \caption{Critical curves and caustics predicted by the different lens models explored in this work, namely ``Shear only'', ``Cluster only'' and our fiducial model ``Cluster+Shear'' that best fits the data. Also shown are the critical lines and caustics from recent models of \elanz, namely from \citet[][constrained by pairs of multiple images with JWST imaging data]{Kamieneski2023} and \citet[][cluster-scale model of \elgor using HST and MUSE data sets]{Caminha2023}. Caustics from our models are shown with dashed lines for clarity. The left panel is focused on the image plane, while the right panel is focused on the source region predicted by models including the cluster mass distribution. Note how the position of the caustics for models that do not include the cluster contribution (i.e., the \citet{Kamieneski2023} and ``Shear only'' models) is aligned with the lens (left panel), compared to the caustics of other models (right panel). The predicted (unlensed) surface brightness of the lens and source galaxies from the best-fit ``Cluster+Shear'' model (red curves) is shown in the background. North and East directions are indicated by the arrows.}
    \label{fig:comparison_mag_lines}
\end{figure*}

We compare in Fig.~\ref{fig:comparison_mag_lines} the critical lines and caustics predicted by our three lens model families. All models predict qualitatively similar critical lines, with a characteristic feature around model position $(0'', 1'')$, on the East of L2. Overall, all models are consistent to each other in the vicinity of the lensed arcs, as expected from the many pixels that contain significant source flux. Nevertheless, we note several features that differ among the models. First, the critical lines from the ``Cluster only'' model do not extend towards L3, which is because this model converges to an Einstein radius $\theta_{\rm E, L3}$ consistent with no mass, despite the prior from the measured stellar velocity dispersion. The critical line resulting from this model also shows a slight deviation outwards, which may be compensating the lack of mass on the east side of L1. Comparing the two models including external shear, we notice slight differences in the critical lines in between L1 and L3 and around L3, which are locations that are less constrained by strong lensing and mostly informed by the prior on the mass of L3 from stellar kinematics. Finally, due to the absence of an explicit cluster mass in the model, the caustics---and thus, the source position---in the ``Shear only'' model largely differ from the two other models. Indeed, the absence of large deflection fields from \elgor leads to the source position being almost aligned with the deflectors, consistent with the observation of strong lensing features. Nevertheless, we note that the absolute source position is generally irrelevant (especially in cluster-scale regime) since numerous unobserved massive structures along the line of sight can shift the  position of the source; hence, only relative positions of source intensity reconstruction relative to caustics are relevant.

In source plane, the tangential caustic have very distinctive features, caused by the combination of individual astroid caustics from the three main deflectors. These features---which are typical of binary or more complex lenses---have been studied in the past with catastrophe theory by describing complex caustic patterns as the product of so-called metamorphoses and critical points \citep[e.g.,][]{SchneiderWeiss1986}. The eastern part of the caustic, towards the left in the right panel of Fig.~\ref{fig:comparison_mag_lines}, is dominated by the astroid caustic from L3. More towards the middle, we see a narrowing of the caustics, which is a consequence of a so-called beak-to-beak metamorphosis \citep[see e.g.,][second row of their Fig.~2]{OrbandeXivryMarshall2009}. The western part of the caustics network corresponds to the merging of L1 and L2 astroid caustics. The southern cusp of these merged caustics is itself composed of two cusps, which is again a typical feature of binary lenses similar to a beak-to-beak transition. According to the classification of \citet{ShinEvans2008}, these merging cusps features can be related to a Type 1 morphology, which in image plane is directly linked to the ``dip'' visible in the northern part of the critical lines around model position $(0'', 1'')$ (see second row of their Fig.~1). The critical lines are also very similar to those shown in Fig.~3 of \citet[][their $m=0$ case]{Bozza2020}. We note that this split cusp structure is visible in all lens models shown in Fig.~\ref{fig:comparison_mag_lines} and, coincidentally, the brightest component of the source galaxy visible in the all NIRCam LW filters lies very close to it in source plane.

\begin{figure*}
    \centering
    \includegraphics[width=0.8\linewidth]{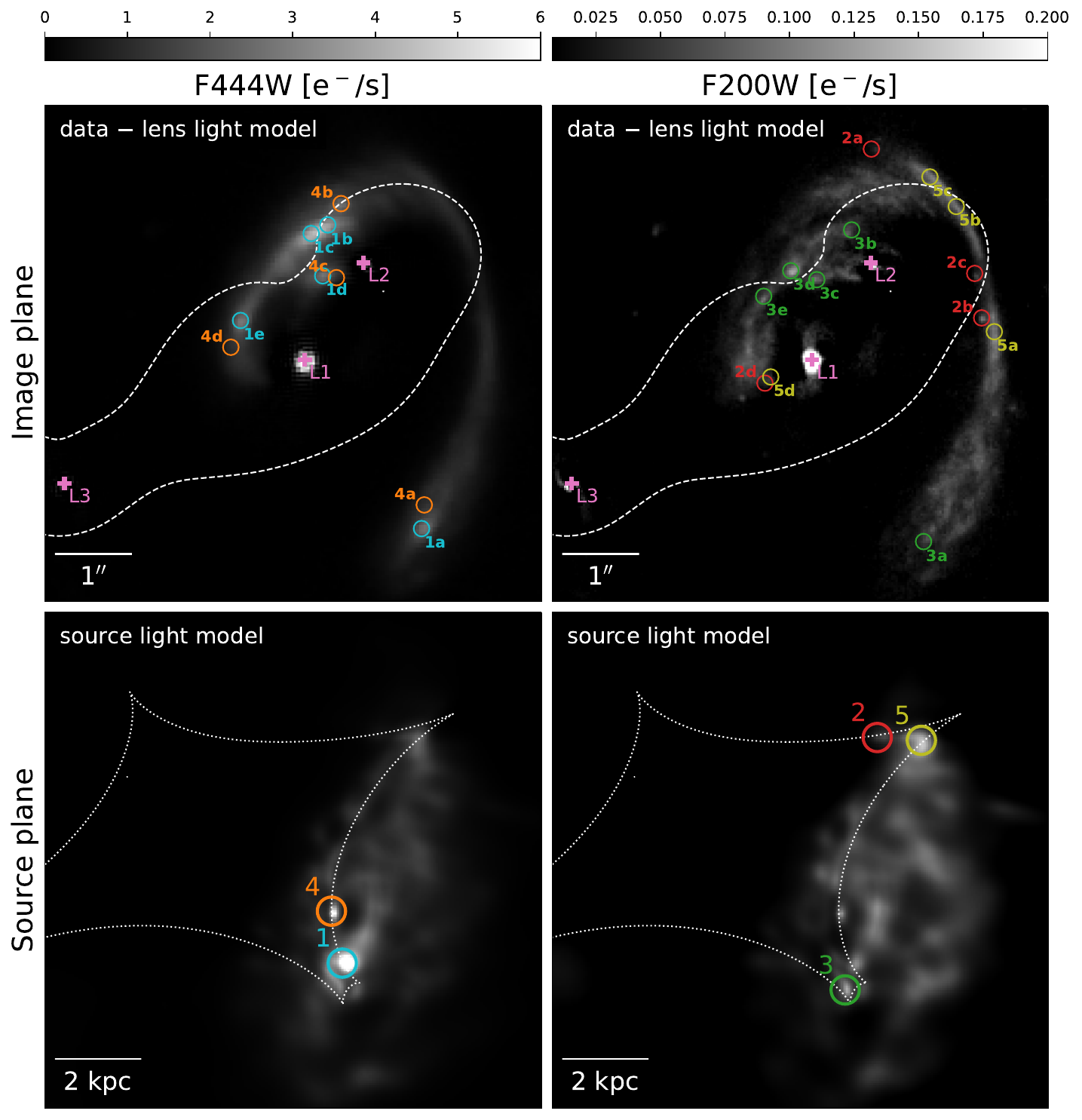}
    \caption{Multiple images of peculiar features in the background source galaxy. \textit{Top row}: families of multiple images in image plane together with the critical lines and the observed flux from the arc (after lens subtraction in filters F444W ad F200W). Also shown are the positions of the three deflectors with pink crosses (residuals remaining after lens light subtraction are visible around L1 and L2 positions), as well as the critical lines with dashed white lines. Our model also predicts images at the position of L1 and L2, although there are not shown here to limit clutter. \textit{Bottom row}: corresponding image positions on the source plane, superimposed onto one sample of our source model. The tangential caustic is shown with dotted white lines. Note that each column has its own dynamic range. Scale bars indicate the angular (in arcsec) and physical (in kpc) scales in image and source plane (assuming $\zs=2.291$), respectively.}
    \label{fig:multiple_images}
\end{figure*}

Although using pairs of multiply imaged features as constraints, the lens model of \citet{Kamieneski2023} is qualitatively very similar to ours. We show their critical lines and caustics\footnote{An offset of $+0\farcs25\ {\rm E}, +0\farcs01\ {\rm N}$ have been added to the critical lines and caustics of \citet{Kamieneski2023} for proper comparison with our models (P. Kamieneski, private communication). We have also found a similar offset between deflectors positions reported in \citet{Caminha2023} and our fitted light profile positions.} in comparison to ours in Fig.~\ref{fig:comparison_mag_lines}. As expected, the main differences in the critical lines are directly related to the amount and location of the multiple images used in \citet{Kamieneski2023}. A striking example is the westernmost part of the arc---around position $(2\farcs2, 1'')$---, for which \citet{Kamieneski2023} used a pair of multiple images which are probably unresolved by JWST and appears as point sources (images 2b and 2c in Fig.~\ref{fig:multiple_images}). In between these two images, all critical lines are overlapping, showing that both the arc and point-like constraints locally give similar results, despite our extended source model not including any point source components. On the contrary, in locations where either pairs of images are missing (e.g., south-western part of the arc) or inaccurate multiple images are used (northern part of the arc), clear differences do arise. In particular, the zoom-in inset in Fig.~\ref{fig:comparison_mag_lines} focuses on a region with many multiple images of central regions of the source galaxy and shows deviations between the critical lines from different models. As we discuss in more detail in Sects.~\ref{ssec:image_families} and \ref{ssec:clump_detection}, we find that one feature in the arc may have been incorrectly attributed to a lensed image of the bright source bulge. We believe that this incorrect attribution causes the deviation between critical lines in the top right corner of the inset in Fig.~\ref{fig:multiple_images}, despite a region with a large number of lensing constraints. Focusing now on the caustics, we see a clear offset in the main astroid caustic between the model of \citet{Kamieneski2023} and our ``Shear only'' model. This result may be surprising at first, since both models are similarly parametrized (power-law profiles embedded in an external shear). However, this offset could be explained by the fact that our models including external shear all predict a non-zero mass for L3, as well as differences in the lensing constraints between the two approaches.

We also compare in Fig.~\ref{fig:comparison_mag_lines} our critical curves and caustics to the ones predicted by the cluster model. The cluster model was constructed without any constraints from \elanz hence L1, L2 and L3 are mainly constrained through scaling relations based on their photometry \citep{Caminha2023}. We find that critical lines between our galaxy-scale model and this cluster-scale overall agree well, in particular in the vicinity L1 and L2. As shown in the inset of Fig.~\ref{fig:comparison_mag_lines}, the characteristic feature in the critical lines is comparable, meaning that qualitatively, the relative masses of L1 and L2 are consistent among the models. The cluster model displays rounder critical lines than our model, as expected from the use of spherically symmetric mass profiles. The discrepancy is the largest in the vicinity of L3, as this galaxy lies further away from the arcs. The cluster mass model assigns a larger mass to L3, which is also visible in source plane, where the eastern astroid caustic is significantly larger than in our models. While investigating this discrepancy is beyond the scope of this work, we simply mention that it may relate to the specific morphology of L3, which appears more elongated than L1 and L2 in the NIRCam data (see also the light models in Fig.~\ref{app:fig:lens_light_model}). In addition, we notice that the light profile of L3 is very steep, as it can be seen from the prominent PSF spikes. Such a surface brightness distribution can be the sign of a disk component superimposed to a bright and compact bulge. In this case, a single-component mass profile like the SIE or dPIE might not be sufficiently accurate \citep[for disk structures in lensing galaxies, see e.g.,][]{Hsueh2018}.

\subsection{Constraints on mass model parameters \label{ssec:mass_constraints}}

\begin{figure*}
    \centering
    \includegraphics[width=\linewidth]{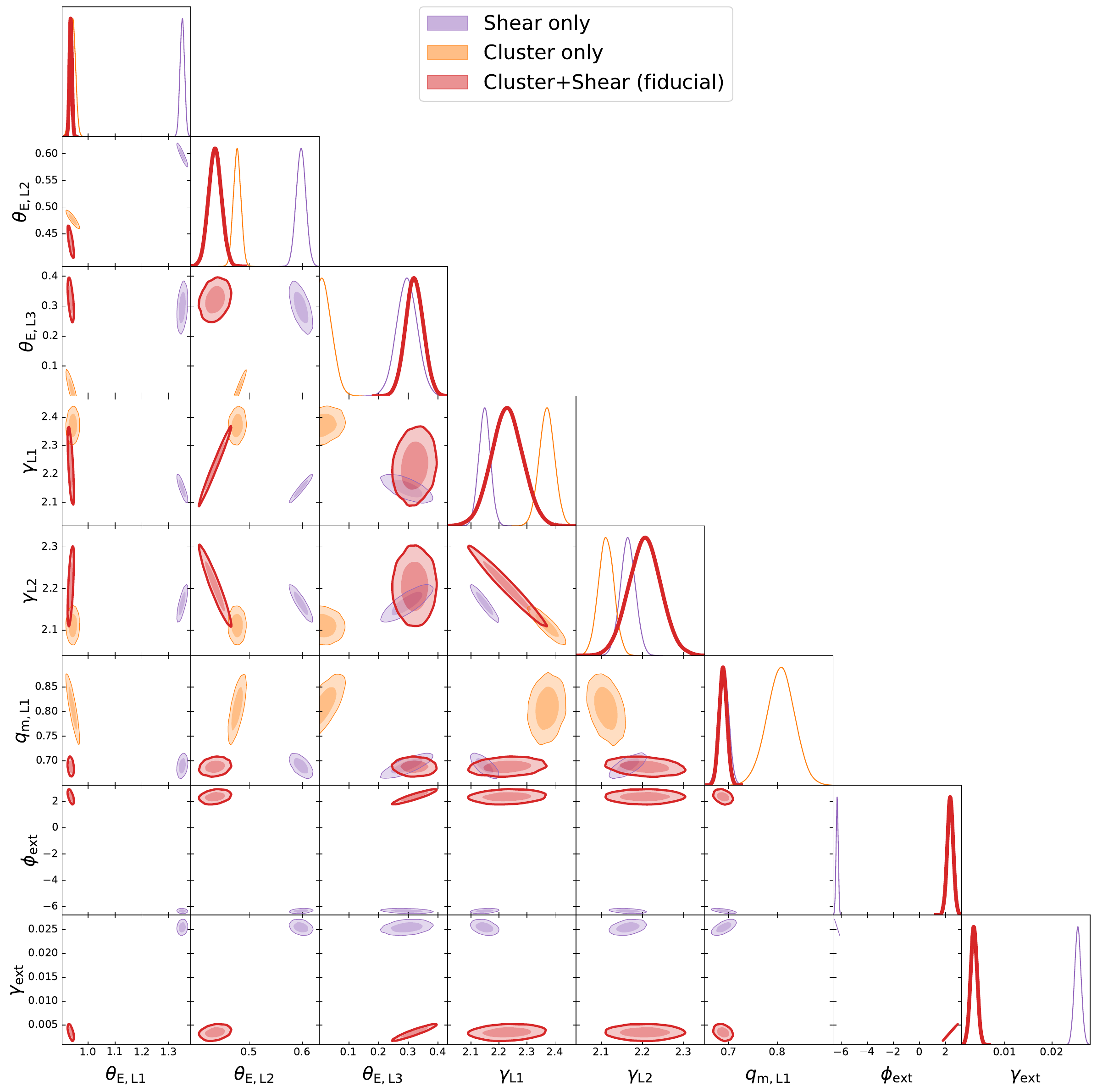}
    \caption{Joint posterior distribution of a subset of lens mass parameters, for the 3 main model families tested in this work. The first one includes only external shear in addition of the three main deflectors, the second one replaces external shear with the \elgor cluster deflection field, and the last one includes a combination of the two. The parameters shown in the figure are the Einstein radii ($\theta_{\rm E, L1}, \theta_{\rm E, L2}, \theta_{\rm E, L3}$), the logarithmic density slopes at $\theta_{\rm E}$ ($\gamma_{\rm L1}, \gamma_{\rm L2}$), the mass axis ratio ($q_{\rm m, L1}$), and the external shear orientation and strength ($\phi_{\rm ext}, \gamma_{\rm ext}$, when included in the model). See Fig.~\ref{app:fig:lens_param_corner_full} for the full mass model parameters space.}
    \label{fig:lens_param_corner_main}
\end{figure*}

\begin{table*}
    \caption{Constraints on the mass distribution of the main deflectors.}
    \label{tab:lens_mass_posterior}
    \renewcommand{\arraystretch}{1.4}
    \centering
    
\begin{tabular}{ccccccc}
\hline\hline
Mass component / Parameter & $\theta_{\rm E}$\ [\,$''\,$] & $\gamma$ & $q_{\rm m}$ (or $\gamma_{\rm ext}$) & $\phi_{\rm m}$ (or $\phi_{\rm ext}$)\ [deg] & $x_{\rm m}$ [\,$''\,$] \tablefootmark{$\dag$} & $y_{\rm m}$ [\,$''\,$] \tablefootmark{$\dag$} \\
\hline
L1 (EPL) & $0.935 \pm 0.005$ & $2.23 \pm 0.05$ & $0.69 \pm 0.01$ & $37.7 \pm 4.7$ & $0.163 \pm 0.004$ & $-0.08 \pm 0.01$ \\
L2 (EPL) & $0.44 \pm 0.01$ & $2.21 \pm 0.04$ & $0.72 \pm 0.01$ & $22.7 \pm 1.6$ & $0.89 \pm 0.01$ & $1.172 \pm 0.004$ \\
L3 (SIE) & $0.32 \pm 0.03$ & $2.00$ \tablefootmark{$\star$} & $0.60$ \tablefootmark{$\star$} & $-0.86$ \tablefootmark{$\star$} & $-2.97$ \tablefootmark{$\star$} & $-1.69$ \tablefootmark{$\star$} \\
External shear & --- & --- & $0.003 \pm 0.001$ & $2.4 \pm 0.2$ & --- & --- \\
Host cluster & \multicolumn{6}{c}{Model from \citet{Caminha2023} without L1, L2, L3} \\
\hline
\end{tabular}

\tablefoot{
The above constraints are computed from the marginalized posterior distributions of our fiducial ``Cluster+Shear'' model. \tablefoottext{$\dag$}{The origin $(0'',0'')$ of our model coordinate system coincides with WCS coordinates $(15.7057851^\circ, -49.2518014^\circ)$, the $x$ axis is positive towards the West and the $y$ axis is positive towards the North.} \tablefoottext{$\star$}{Except for $\theta_{\rm E}$, all L3 parameters are fixed to an isothermal slope ($\gamma=2$) or to values informed by the surface brightness model.}
}
\end{table*}

In this section we present our quantitative constraints on the mass distribution of the main deflectors. We show in Fig.~\ref{fig:lens_param_corner_main} the joint posterior distributions of a subset of key mass model parameters, for each of the three model families (see Appendix~\ref{app:ssec:lens_mass_full_post} for the full corner plot). We also summarize in Table~\ref{tab:lens_mass_posterior} all constraints on mass model parameters for the fiducial ``Cluster+shear'' model. As expected, the Einstein radius of L1 and L2 as well the external shear strength significantly decrease after including the cluster mass distribution. Unlike the model of \citet{Kamieneski2023} composed of SIE profiles embedded in an external shear (which is the most similar to our ``Shear only'' model) the mass of L3 is constrained by the imaging data, except for the ``Cluster only'' model family. Our fiducial model predicts $\theta_{\rm E,L3}=0\farcs32\pm0\farcs03$, which is $\sim3\sigma$ away from the stellar kinematics estimate $\theta_{\rm E,SIS}=0\farcs49\pm0\farcs05$. Therefore, the prior from stellar kinematics appears rather unconstraining compared to the imaging data, despite the angular distance between L3 and the lensing features.

In terms of the radial mass density profiles of L1 and L2, all model families predict logarithmic density slope values steeper than isothermal. Our fiducial ``Cluster+Shear'' model predicts density slope values of $\gamma_{\rm L1} = 2.23\pm0.05$ and $\gamma_{\rm L2} = 2.21\pm0.04$. These values are inconsistent with an isothermal slope at the $\sim5\sigma$ significance level. We also note the anti-correlation between the density slopes of the two deflectors from the joint $\gamma_{\rm L1}-\gamma_{\rm L2}$ posterior distribution in Fig.~\ref{fig:lens_param_corner_main}. In Sect.~\ref{ssec:density_profiles_discussion} we discuss further these density slope constraints and place them in the broader context of galaxy evolution.

\begin{figure}
    \centering
    \includegraphics[width=\linewidth]{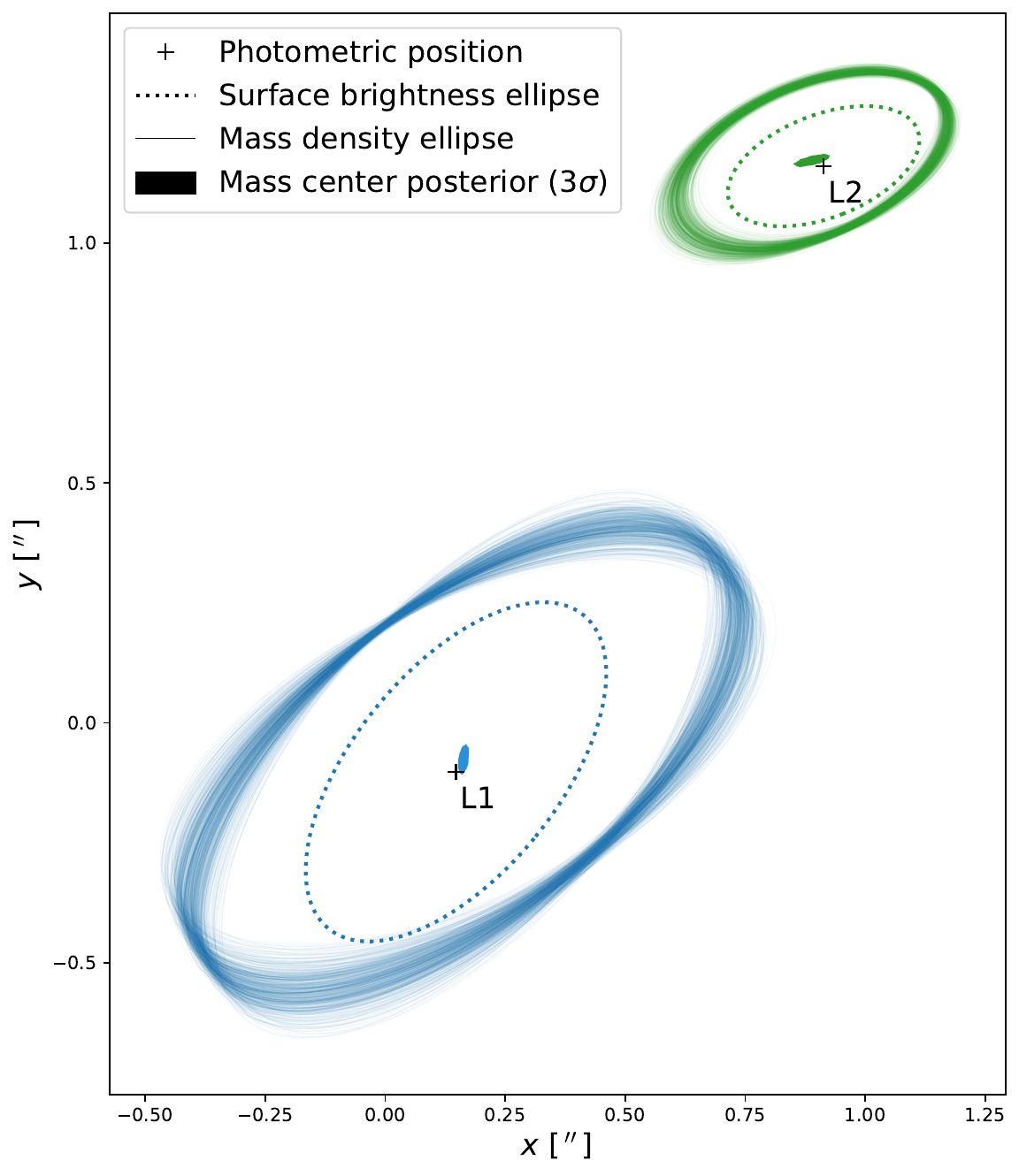}
    \caption{Visualization of position, ellipticity and angular size of the deflectors L1 and L2. For each galaxy, the cross shows the mean position across the six bands from the fit to the surface brightness (double-Sérsic profile), and the dotted line shows an ellipse to visualize that fit whose radius corresponds to the half-light radius of the profile. The sets of thin lines are ellipses corresponding to 500 individual samples of the lensing convergence (EPL profile) with radius equal to the Einstein radius, drawn from the posterior distribution of Fig.~\ref{fig:lens_param_corner_main}. Close to the position of each deflector, small shaded contours indicate the posterior distribution of the EPL center (extending to $3\sigma$). For each deflectors, we see offsets in position angle and centroids between their mass and light distributions.}
    \label{fig:lens_ellipses}
\end{figure}

We compare our constraints on the mass distribution to those on the lens light distribution in order to detect possible offsets between the two components. We do so by showing in Fig.~\ref{fig:lens_ellipses} a simplified view of the morphology of L1 and L2. The mass ellipticity of L1 is slightly larger than its luminous counterpart. While the orientations of the light and mass distributions of L2 are similar, those of L1 are slightly offset from each other, and the surface brightness of L1 seems to be oriented towards L2 more than its mass density. From Fig.~\ref{fig:lens_ellipses} we also notice that the position angle of L1 is less constrained than for L2. This is not entirely surprising since the latter appears closer to the lensing features. Moreover, we find that the mass centroids of L1 and L2 are offset by about $3\sigma$ from their corresponding light centroids. More specifically, the center of mass of L1 is offset towards L2, while the one of L2 is offset towards the North-East. These offsets may be attributed to signs of gravitational interaction between the two galaxies. These differences in axis ratio and position angles are also visible in the full posterior distribution shown in Fig.~\ref{app:fig:lens_param_corner_full}.

In Appendix~\ref{app:ssec:systematics} we extensively check the robustness of our constraints against different modeling choices. These checks include the choice of VI sampler, the effect of inaccurate lens light model and the uncertainty associated to the cluster-scale model. In particular, we note that our models very weakly dependent on the source resolution, both in terms of constraints on mass model parameters and morphology of the reconstructed source. We believe this low dependence on the source resolution is a direct consequence of the efficient regularization of our correlated field source model, which allows us to reach high-resolution without compromising on computation time and model quality.

Finally, we find our lens model to be sensitive to changes in the cluster-scale mass distribution, as drawing different posterior samples from the cluster model of \citet{Caminha2023} lead to measurable changes in $\chi^2_\nu$. In particular, some posterior samples from the cluster model lead to a lower $\chi^2_\nu$ than with the best-fit cluster model we use in our fiducial model. Therefore, \elanz lensing features may be used to discriminate between different cluster models, thus providing additional constraints on the mass distribution of large-scale components of \elgor (i.e., its dark matter halo). Although this particular result demands further investigation, it shows that extended arcs, although primarily constraining the structure of individual cluster members, may also provide useful constraints on the mass distribution at the scale of the cluster.

\subsection{Lensed image families and clumps in the source \label{ssec:image_families}}

\begin{figure}
    \centering
    \includegraphics[width=\linewidth]{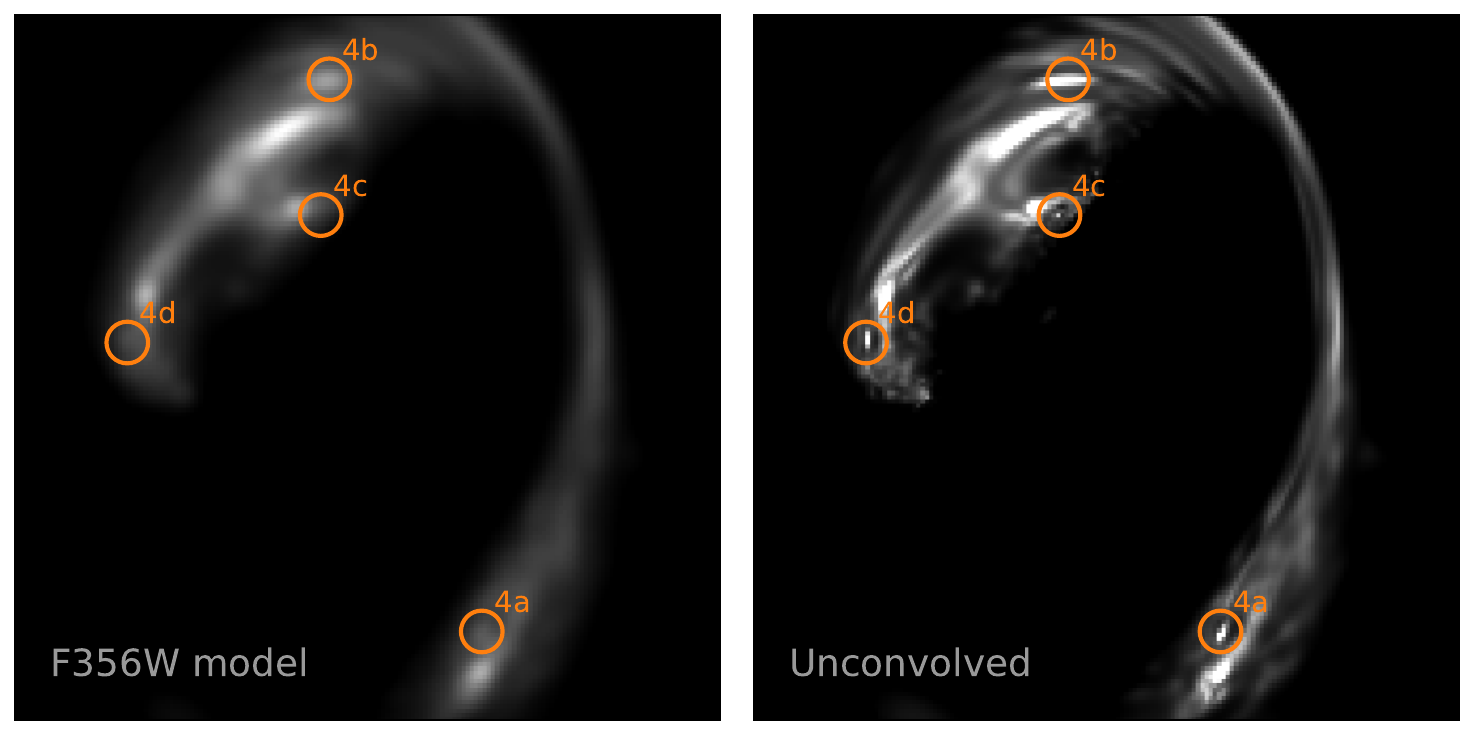}
    \caption{Multiple images of a $\lesssim400$ pc compact clump in the source galaxy (image 4 in Fig.~\ref{fig:multiple_images} and shown in color in Fig.~\ref{fig:model_color}). Only image 4a is clearly visible in the data and our model, shown in the left panel (shown here for filter F356W, which also looks similar in F277W, F410M and F444W filters that are not shown). The right panel shows for comparison the model before convolution by the PSF, which reveals the four multiple images (the colormap slightly saturates the brighter pixels to improve contrast).}
    \label{fig:zoom_source_substructure}
\end{figure}

As the \elanz lens system is composed of multiple deflectors and is directly influenced by its host cluster, it is not straightforward to assign a lensing origin to each individual feature observed in the arc. We use our fiducial lens model to show in Fig.~\ref{fig:multiple_images} a set of multiply lensed images that correspond to prominent features either visible in the observed arcs or in the unlensed source model. Within a given image family, the lensed images are sorted according by increasing Fermat potential, or equivalently by their arrival time \citep{Schneider1985}. For image families that were previously mentioned in \citet{Kamieneski2023}, we try to follow as closely as possible their original numbering. This is the case for our image families 1 (left column of Fig.~\ref{fig:multiple_images}), and 2 and 3 (right column). However, since we provide more multiple images per image family---and we find that image 1b from \citet{Kamieneski2023} does not belong to family 1---the specific letter used in image identifiers must differ from theirs. \citet{Diego2023} also refers to three families of multiple images, although without explicit identifiers (see their Fig.~2). We note that, for all image families shown in Fig.~\ref{fig:multiple_images}, the lens equation also predicts images appearing at the position of L1 and L2. These are extremely demagnified images (given the cuspy nature of power-law profiles with super-isothermal slope). We do not show these central images in order to avoid cluttering the figure.

We are able to confirm that parts of the background galaxy are multiply imaged five times (excluding central images); in Fig.~\ref{fig:multiple_images} this is shown by image families 1 and 3. This peculiar lensing configuration has been first suspected in \citet{Kamieneski2023}. We check if these fifth images are magnified by computing their model-predicted magnification from our fiducial mass model. We obtain magnifications of $\mu \approx -3$ for image 1d and $\mu \approx -5$ for image 3b, thus we find there are magnified. We note that this ``exotic'' lensing configuration typically arises in systems with multiple deflectors and has been the topic of several theoretical works in the past \citep[e.g.,][]{ShinEvans2008,OrbandeXivryMarshall2009,Bozza2020}. We briefly discuss in Appendix \ref{app:ssec:central_images} potential implications regarding this lensing configuration.

\subsection{Detection of multiply imaged clumps in the source \label{ssec:clump_detection}}

We find that one of the images used in \citet{Kamieneski2023} as being part of the image family 1 (their image 1b) in fact does not belong to this set of multiple image according to our lens model. On the one hand, we find that their image 1d is in fact composed of the two images denoted by 1a and 1c, which appear like merging while crossing the tangential critical line. On the other hand, their image 1b corresponds to our image 4a in Fig.~\ref{fig:multiple_images}, and thus originates from a distinct compact bright clump in the source, located at the north of the central bulge. We note that this clump has similar colors as the bulge of the source galaxy, which is why it could have been incorrectly attributed to a lensed image of the bulge. Interestingly, not all images of this clump are visible in the data. We believe it is mostly due to a combination of a the larger FWHM of the PSF in redder bands, and stronger contamination by the deflectors light. To better visualize this, we show in Fig.~\ref{fig:zoom_source_substructure} a comparison between our the lensed and convolved source model and the corresponding model before convolution by the PSF (in F356W although it is similar in other filters). We see that only image 4a remains clearly visible after convolution. Moreover, the lensed images appear either close to the bright bulge of the source (image 4d), or close to regions heavily contaminated by the deflectors light (images 4b and 4c); therefore, they are not seen in the data. From one of our higher resolution source models (with $n_{\rm src}=300$), we estimate that the physical diameter (in projection) of the clump is approximately $400$ pc at redshift $\zs=2.291$. We measure that this clump is located at approximately $1.2$ kpc from the source centroid. While the size estimate is an upper-bound to the true size---since increasing the source resolution tends to make it more compact---this clump in the source is remarkably small. It is comparable in size to some of the structures in the jets of lensed sources analyzed in \citet{Powell2022} using VLBI observations with milli-arcsecond resolution (against 30 milli-arcsecond resolution with JWST/NIRCam data).

Finally, we note that our extended source model can be used to extract the SED of compact unresolved structures in the arc. We show in Appendix~\ref{app:sec:clump_sed} the case of one of the brightest point-like feature in \elanz (image 2b in Fig.~\ref{fig:multiple_images}). As our pixelated source model, at its fiducial resolution ($n_{\rm src}=100$), does not capture unresolved features by construction, it can be directly subtracted from the data to separate the extended source component from these unresolved features. It is then easier to perform photometric measurements and extract the SED of such compact source regions, which are important tracers of the star formation history of the galaxy \citep[as shown in][]{Kamieneski2023}.

\subsection{Summary of the main findings}

One of the main results of our strong lensing analysis of \elanz is the measurement of mass density slopes of the cluster member galaxies L1 and L2 at redshift $z\approx0.9$. We find that both deflectors, especially the lower mass galaxy L2, have density profiles steeper than isothermal from $3\sigma$ to $9\sigma$ statistical significance. These constraints are obtained by modeling the full surface brightness of the lensed dusty galaxy, for which we detect and reconstruct a small luminous clump of size $\lesssim400$ pc at redshift $z\approx2.3$. We discuss in more details these findings and place them in a broader context in Sect.~\ref{sec:discussion}.

\subsection{Release of lens models within the COOLEST standard}

Previous sections describe our modeling and inference approach with a level of details that aim at maximizing the reproducibility of our analysis. With this idea in mind, we also publicly release all lens models\footnote{\url{https://github.com/aymgal/ElAnzuelo_modeling_public}} following the strong lensing standard COOLEST \citep{Galan2023_coolest}. These models can then be easily loaded using the Python package \texttt{coolest}\footnote{\url{https://github.com/aymgal/COOLEST}} for further analyses, in particular with any modeling codes that share an interface with COOLEST (see the documentation for a list). We encourage future lens models to be released under this standard, in order to facilitate comparison with the present analysis.

\section{Discussion \label{sec:discussion}}

\subsection{Mass density profile of elliptical galaxies \label{ssec:density_profiles_discussion}}

The three foreground deflectors if \elanz have typical photometric signatures of elliptical galaxies and have been confirmed to be members of the massive cluster \elgor \citep[e.g.,][]{Caminha2023}. In this work, we model the mass distribution of these galaxies using elliptical power-law profiles perturbed by the cluster mass distribution and an additional external shear. As presented in Sect.~\ref{sec:results} (Table~\ref{tab:lens_mass_posterior}), we infer logarithmic density slopes for L1 and L2 of $\gamma_{\rm L1}=2.23 \pm 0.05$ and $\gamma_{\rm L2}=2.21 \pm 0.04$, respectively. These values are inconsistent with an isothermal slope, and $\gamma_{\rm L2}$ is $2.7\sigma$ away from the mean value $\langle\gamma\rangle=2.075\pm0.024$ of \citet{Etherington2023_profiles} inferred from a subset of SLACS \citep{Bolton2006} and GALLERY \citep{Shu2016} systems. As detailed in Sect.~\ref{sec:image_modeling}, we infer our uncertainties by marginalizing over an ensemble of model variations with Bayesian weighting. We also looked for potential systematic effects and did not find statistically significant changes in the posteriors distributions by running additional models. Here, we first compare the precision of these measurements with previous analyses of similar imaging data, then discuss the possible origins of these results in terms of galaxy structure.

Two galaxy-scale strong lenses observed in multiple NIRCam filters have been analyzed at the time of writing this manuscript: the backlit-galaxy system VV\,191 from the PEARLS program \citep{Keel2023} and the dusty strongly lensed galaxy SPT0418$-$47 from the TEMPLATES program \citep{Cathey2023,Rustig2024_lenscharm}. For both of these systems however, the data quality and lensing features are not comparable to the PEARLS observations of \elanz (almost no counterimage for VV\,191, small separation and low S/N for SPT0418$-$47). Moreover, the density slope parameter has been fixed to the isothermal value in both analyses. Therefore we turn to analyses of multi-band imaging data from HST instead, acknowledging that those are usually based on a smaller number of filters. One example is the semi-automated modeling effort of \citet{Tan2023}, where the authors jointly fit between 1 and 3 HST filters\footnote{The system SDSS\,J$0912+0029$ is the only system modeled using 4 filters, but is excluded from their analysis due to high lens light contamination and low S/N for the arcs.} in the visible and ultraviolet wavelengths for a selection of SLACS and SL2S systems. The average uncertainty on the density slope is $0.03$ (without their additional systematic error term and for SLACS systems only, as those have overall higher S/N than SL2S) which, reported to our mean value for L1 gives to a relative precision of $1.4\%$. This relative precision is smaller than ours ($2.2\%$ and $1.8\%$ for L1 and L2, respectively), despite our lens model being constrained by 6 filters at higher resolution. This gives further confidence that our uncertainties on $\gamma$ are not under-estimated, since they are larger than those typically obtained with typical HST strong lensing data.

That elliptical galaxies have nearly isothermal radial density profiles in their inner region has been observed in isolated deflectors, such as those from the SLACS \citep{Bolton2006} and SL2S \citep{Gavazzi2012} samples. The three main deflectors of \elanz are, however, not isolated but genuine members of \elgor that are gravitationally interacting. Therefore, it may not be surprising to measure density slopes that deviate from an isothermal profile. For most of isolated lens systems from SLACS (or comparable samples) we note that the ratio $\bar{r}_{\rm E/eff}$ between the Einstein radius and the half-light radius, $\bar{r}_{\rm E/eff}\equiv\theta_{\rm E}/\theta_{\rm eff} \lesssim 1$, meaning that lensing overall constrains the inner region of the lens galaxies, within their half-light radius \citep[e.g., see the middle panel of Fig.~15 from ][for SLACS and GALLERY samples]{Etherington2022_pyauto}. For the case of \elanz, L1 and L2 both have $\bar{r}_{\rm E/eff, L1} \approx 1.6$ and $\bar{r}_{\rm E/eff, L2} \approx 1.3$ respectively (as visualized in Fig.~\ref{fig:lens_ellipses}), meaning that lensing provides constraints at overall larger radii compared to lens galaxies from the aforementioned samples\footnote{The effective radius for L2 may be underestimated due to due blending with the lensed arcs (see Appendix~\ref{app:ssec:lens_light}). However, due to the lensing boost by the cluster, the ratio $\bar{r}_{\rm E/eff,\rm L2}$ is likely larger than unity.}. The main reason for \elanz deflectors to have $\bar{r} > 1$ is that their apparent lensing effect is boosted by the mass density of the host cluster, leading to the appearance of an Einstein ring that is larger than if L1 or L2 were isolated galaxies. This effect can be qualitatively confirmed by noting the difference in $\theta_{\rm E}$ values between the ``Shear only'' and ``Cluster+Shear'' models (e.g., Fig.~\ref{fig:lens_param_corner_main}): including the cluster mass distribution in the model significantly reduces \thetaE values for L1 and L2 in order to fit the imaging data. Therefore, the observed lensing features probe the mass distribution of the deflectors at sensibly larger radii, compared to isolated deflectors.

Beside larger Einstein radii in the vicinity of their host cluster, it has been shown that cluster members are expected to have truncated radial density profiles, in particular in the low to intermediate mass range \citep[e.g.,][]{Natarajan2002,Monna2015}. Truncated mass distributions typically arise in cluster environments as a result of tidal stripping. Recently, \citet{Granata2023} discuss opportunities and challenges that exist when constraining the truncation radius $r_{\rm t}$ and central velocity dispersion $\sigma$ parameters of cluster members. In the case of \elanz, it is interesting to estimate a range of plausible truncation radii expected for deflectors L1 and L2. We can do so by looking at the $r_{\rm t}-\sigma$ relation summarized in Fig.~11 of \citet{Monna2015}. For a velocity dispersion $\sigma_{\rm L1}\approx300$ \ks for L1 ($\sigma_{\rm L2}\approx200$ \ks for L2), the truncation radius is expected to fall in the range $10-100$ kpc ($8-70$ kpc). The circularized Einstein radius in physical units $r_{\rm E}$ inferred from our lens model is $r_{\rm E,L1}\approx7$ kpc ($r_{\rm E, L2}\approx3$ kpc), likely smaller than the truncation radius. However, the asymmetric arcs of \elanz extend to a maximal projected distance of $\sim40$ kpc for L1 ($\sim15$ kpc for L2), that is within the range of plausible truncation radius values. Therefore, it may be possible to constrain the truncated profile of these cluster members from extended source modeling, especially for L2 due to its smaller size. In the same study, \citet{Granata2023} also mention that $r_{\rm t}$ and $\sigma$ are degenerate \citep[see also][]{Bergamini2019}, and discuss how different methods and observables help mitigating this issue. In particular, they make the parallel between two types of lensing analyses: (1) point-like multiple images appearing close to cluster members over a possibly large range of radii \citep[as in][]{Granata2023}, and (2) single extended arcs warping around cluster members \citep[e.g., as in][]{SuyuHalkola2010,Monna2015}. The particular case of \elanz seems to bridge the gap between the two situations, as lensing features have both high S/N and a significantly assymetric shape, such that they cover a large range of galactic radii. In summary, \elanz is a promising system to go beyond simple density profiles and possibly characterize tidal truncation, as suggested by the steep density profiles we infer.

Finally, we note that measurements of a single mass density slope from lensing may be affected by systematic biases caused by simplistic modeling assumptions \citep[e.g.,][]{Sonnenfeld2018a,Kochanek2020,Cao2022}. If the intrinsic radial profiles of the deflectors are more complex (e.g., with one or more inflection points), approximating those with a single power-law profiles can result in a biased interpretation of the constraints, either over- or under-estimating the slope depending on the location of the lensing features \citep[e.g.,][]{GomerWilliams2020,Millon2020_TDC_I}. In the case of \elanz, the considerable thickness of the arcs and the very asymmetric lensing configuration---the northern arcs being significantly closer to L1 and L2 in projection---may provide constraints for a more flexible model, along both radial and azimuthal directions.

\subsection{Approximating the host cluster with an external shear}

\begin{figure}
    \centering
    \includegraphics[width=\linewidth]{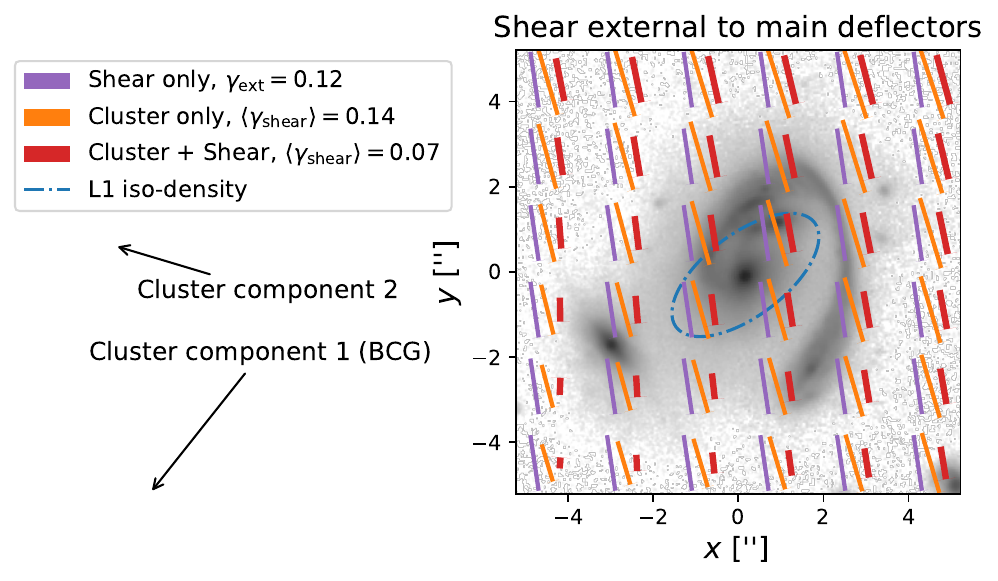}
    \caption{Shear orientation and strength over the field of view of \elanz. For the ``Shear only'' model, the shear field is a uniform external shear with strength $\gamma_{\rm ext}$; for the ``Cluster only'', it is the shear field from the cluster model of \citet{Caminha2023}; for the ``Cluster+Shear'' model, it is the combined effect from the uniform external shear and the cluster shear. For the latter two models, the averaged shear strength $\langle\gamma_{\rm shear}\rangle$ is also indicated in the legend. The absolute length of the shear markers is arbitrary, but their relative size is respected. The blue ellipse indicates the orientation of the mass density of L1. On the left of the plot, we show the directions pointing towards the two main dark matter halo component of the \elgor cluster. Note that the shear orientation is consistent with the position of the cluster scale components.}
    \label{fig:comparison_shear_fields}
\end{figure}

In most of galaxy-scale strong lensing analyses, the net effect of masses external to the main deflectors is usually approximated by a constant external shear field. Here we investigate further the assumption of a constant external shear, compared to the explicit spatially varying contribution of the cluster. To help visualizing the various shear fields at play, we compare the best-fit shear fields (that are external to L1, L2, L3) from our three model families in Fig.~\ref{fig:comparison_shear_fields}.

We are interested to verify if the host cluster \elgor, at the position of \elanz, could be modeled using an approximation. As reported in Sect.~\ref{sec:results}, we find that external shear alone leads to a better fit to the data compared to using the fixed deflection field of the cluster, with a $\chi^2_\nu$ difference of $>0.02$. This results may seem surprising: a constant external shear field---which remains a first-order approximation of the cluster's true shear field---leads to a better fit than the full cluster model. In other words, an optimizeable two-parameters approximation performs better than the more complex but fixed model. We argue that these result may in fact be expected: the cluster mass model is the most accurate in the vicinity of multiple image pairs used as constraints. Among the families of multiple images used to constrain the cluster-scale mass model, the closest images are located at approximately $15''$ North-East from \elanz \citep[images 3b and 17c in Fig.~1 of][]{Caminha2023}. Further away from these constraints, the extrapolated model may be less accurate compared to the constraining power of the arc of \elanz. On the contrary, the external shear with free parameters can adjust to this constraining data and lead to a better fit.

Our fiducial model combines the fixed cluster field with local corrections from an uniform external shear. We visualize the different external shear fields involved in Fig.~\ref{fig:comparison_shear_fields}, which also indicates directions towards the two cluster-scale dark matter halos of \elgor \citep{Caminha2023}. We see that the shear from the cluster is mainly mainly driven by the position of the BCG. The uniform correction by the external shear has the net effect of re-orienting the shear direction more towards the second main component of \elgor, which is located towards to the North-East of \elanz. The fact that constraints from \elanz allows us to locally correct the cluster-scale model is another hint that extended arcs are complementary to cluster-scale lensing observables.

\subsection{Using extended arcs instead of point-like constraints}

In this work we use the full information encoded in the arcs in multiple wavelengths to constrain the mass distribution of the deflectors. Compared to only using families of point-like images, pixel-level modeling has two main advantages: the risk of mistakenly assigning pairs of images that do not share the same physical origin is minimized, and the number of lensing constraints is maximized. In Sect.~\ref{ssec:image_families}, we selected and labeled multiple compact regions in source plane and computed their corresponding multiple lensed positions. Similar to the concept of geometric redshifts \citep[see e.g.,][]{Diego2023}, we argue that our lens model is better at securing pairs of multiple images in the absence of a spectroscopic confirmation, compared to relying solely on color and morphological information. Strengthening this argument is the remarkable agreement between our model and the model of \citet{Kamieneski2023}. Their model was built using point-like constraints such as the unresolved pair of images that we name 2b and 2c in Fig.~\ref{fig:multiple_images}. Our model predicts critical lines that pass in between the two images, despite the feature being absent in our source model, meaning that the extended arcs are at least as good as these two unresolved features to constrain the lens model at this location.

In Sect.~\ref{ssec:image_families} we compared in more details our results with the lens model of \citet{Kamieneski2023}, especially for three image families. In their analysis, the authors used the 8 NIRCam filters to look for compact features in image plane that have consistent colors, and considered those as multiple images of same region in the source. For their image family 1, they used a simplified initial lens model to confirm the five-image configuration, interpreted as a confirmation of the origin of images 1a to 1e being lensed versions of the bright central region of the source galaxy. In Sect.~\ref{ssec:clump_detection}, we show that image 1b does not originate from the center of the source, but rather from an off-centered compact clump of size $\lesssim 400$ pc. In NIRCam images, this clump has similar colors as the central bulge of the source \citep[see Fig.~2 of][]{Kamieneski2023}, which is the reason why the only visible image of the clump has been associated to the bulge in source plane. Consequently, we caution on using solely photometric signatures (shapes or colors) to draw conclusions on the lensing origin of features in the arcs. The use of incorrect pairs of multiple images is likely to bias the inference on the source morphology and mass model parameters. However, depending on the application and the required precision of the measurements, these biases may not always affect the conclusions of the analyses. In particular, the analysis of \citet{Kamieneski2023} is likely unaffected, as it focuses on large-scale properties of the source, such as the stellar mass, star formation rate and color gradients over kiloparsec scales.

Our analysis makes use of all pixels that contain lensing features in the six NIRCam filters we selected. Another approach, which can be seen as intermediate between point-like and pixel-level modeling, is to locally extract shape information from individual multiple images to constrain local properties of the lens potential. This idea has been formalized and explored in several works  \citep{Wagner2016,Tessore2017,Fleury2019,Birrer2021}, although with limited applications to real observations so far \citep[e.g.,][]{Wagner2018}. The main advantage in locally constraining the lens potential is to release most of the assumptions regarding the global lens mass distribution, at the cost of additional assumptions regarding the symmetry and compactness of the source (or individual features within it). In systems such as \elanz, the lensed source is likely too asymmetric to directly apply such a formalism. However, it may be possible to combine constraints from extended source modeling with macro-model independent corrections based on the shape of individual lensed resolved clumps within the arc \citep[see e.g.,][for an automated multi-band extraction of lensed features]{Lin2023}.

\subsection{Limitations and future improvements \label{ssec:limitations}}

While we present the most advanced model of \elanz to date, our analysis relies on a set simplifying assumptions. Our strongest assumption is the modeling of the light and mass distributions with smooth elliptical profiles. The model residuals (see Fig.~\ref{fig:bestfit_lens_model_summary}) reveal that the lens light distribution is more complex than what can be modeled with concentric Sérsic profiles, with possible signs of boxyness and tidal stripping due to the interaction between the three deflectors. If already visible in the light distribution, the underlying mass distribution is likely to be more complex than single elliptical power-law profiles, in particular requiring additional degrees of freedom in the radial and azimuthal directions. First-order deviations from the power-law profile can be added, for example, through multipoles (typically of order 3 or 4 \citep[e.g.,][]{VanDeVyvere2022,Galan2022,Powell2022}, cored and truncated profiles \citep[e.g.,][]{Eliasdottir2007} or broken power-law profiles \citep[e.g.,][]{Du2020}. Complexity in the lens light could be added similar to our source model, or using similarly flexible methods like sparsity and wavelet regularization \citep{Galan2021}. Jointly fitting the lens light with the other model components would add extra constraints to lens mass distribution, similar to so-called composite lens models in which the baryonic and dark matter components are modeled as separated profiles \citep[as commonly used in lensed quasar cosmographic analyses,][]{Suyu2014,Millon2020_TDC_I}. We defer the use of more elaborate lens models to future works.

Regarding spectroscopic data, in this work we only attempt to extract the central velocity dispersion from the VLT/MUSE observations. Nevertheless it may be possible to measure the velocity dispersion at various locations around the center of L1 and L3, and perhaps properly separate the flux of L2 from L1. Additional measurements points could help constraining further the mass distribution of the deflectors, in particular L3 for which we find that a single velocity dispersion is less constraining that the lensing data. Spatially resolved kinematics measurements can also help mitigating degeneracies inherent to lens models \citep[e.g., the mass-sheet degeneracy,][]{TreuKoopmans2002,Schneider2013,Gomer2023,Shajib2023_TDC_XII}. Using stellar kinematics of the lensed galaxy can also mitigate model degeneracies, in addition to provide insights on the dynamical properties of \elanz \citep{Rizzo2018,Chirivi2020}. However, extracting reliable measurements at the location of the arcs would require much deeper exposures than the present MUSE data.

Finally, we draw attention to the limited knowledge of noise characteristics in JWST data. Based on our model of \elanz, we notice that noise levels seem over-estimated (alternatively, the exposure time under-estimated) in some locations. One can see in the central column of Fig.~\ref{fig:bestfit_lens_model_summary} that normalized residuals, in some regions within the arc mask, appear ``flatter'' than expected from pure random noise. This local over-estimation of the noise level may lead in turn to an over-estimation of parameters posterior uncertainties. Although for this reason we believe our main results are unaffected, we emphasize that it will be important to better characterize the noise statistics when modeling JWST/NIRCam pixels (not only for lens modeling analyses), in order to get reliable parameters uncertainties\footnote{Close to completion of this work, error maps (``\texttt{\_err}'' files) have been uploaded on the PEARLS website. We compared best-fit models and corresponding $\chi^2$ values obtained using either our empirical uncertainty estimation or these error maps, and noticed that the latter lead to $\chi^2_\nu$ values significantly lower than 1, especially in SW filters. Therefore, we choose to keep our noise maps based on estimated exposure maps.}. As shown by the PEARLS \citep{Windhorst2023_PEARLS} and TEMPLATES \citep{Rigby2023_TEMPLATES} programs, special care in the reduction of JWST spectroscopic and imaging data must be taken, in particular the challenging removal of 1/$f$ noise patterns. These complex reduction steps make it challenging to fully control and understand the noise properties. We encourage future works to explore the impact of the noise properties on lens modeling analyses.

\section{Summary and conclusion \label{sec:conclusion}}

In this work, we model in details the prominent strong lens system \elanz in the vicinity of the massive galaxy cluster \elgor. We do so using high-resolution and multi-band JWST/NIRCam imaging data obtained from the PEARLS program \citep{Windhorst2023_PEARLS} and MUSE/VLT specstroscopic data \citep{Caminha2023}. This system is composed of three elliptical galaxies as main deflectors (L1, L2, L3) at redshift $z_{\rm d} \approx 0.9$ and a strongly lensed dusty star-forming galaxy at redshift $\zs \approx 2.3$. We summarize below the novelty of our analysis and its main results:
\begin{itemize}
\setlength\itemsep{0.8em}

    \item We apply, for the first time on real data, the differentiable lens modeling code \herculens \citep{Galan2022} to model the large number of NIRCam pixels. We combine \herculens with a novel three-dimensional source model and a variational inference approach using \textsc{NIFTy.re} \citep{niftyre}. Our inference pipeline can run on both CPUs and GPUs, the latter allowing us to obtain approximate posterior distributions in about 2 hours, for $\sim10^5$ model parameters constrained by $\sim10^6$ data points.
    
    \item We measure the line-of-sight stellar velocity dispersion for two of the deflectors using VLT/MUSE data, and use these measurements to further constrain the mass of deflector L3. However, we find that the current precision on the velocity dispersion only marginally impacts the constraints from the lens models.
    
    \item We extensively compare our lens model based on extended source modeling to the analysis of \citet{Kamieneski2023} based on point-like constraints within the arc. We show a good qualitative agreement between the models, confirm the five-image lensing configuration and demonstrate how extended source modeling can be used to secure the nature of compact features in the lensed arcs. We caution on relying solely on color and morphology information, since it can lead to incorrect identifications of multiple images.
    
    \item We find that cluster members L1 and L2, which create most of the strong lensing effect, have density slopes at Einstein radius steeper than isothermal with $\gamma\gtrsim2.2$. We make the hypothesis that such steep density slopes are caused by two factors. (1) The lensing boost from the host cluster increases the Einstein radius, such that lensing features probe further away from the deflectors centroids. (2) The mass density profile of cluster members is tidally truncated, such that their measured density slopes can appear super-isothermal.
    
    \item We show that a uniform external shear field does not accurately approximate the cluster mass distribution over the field of view of \elanz. Our best-fit lens model explicitly incorporates the cluster lensing contribution from the model of \citet{Caminha2023}, and uses the external shear component to locally correct its deflection field over the field of view.
    
    \item Our high-resolution multi-band source model allows us to detect and locate a small luminous clump located at $1.2$ kpc from the source centroid and with a maxmimal diameter of $400$ pc at the source redshift. To our knowledge, this may be one of the first clumps of this size located in a dusty galaxy at redshift $z\sim2.3$. Only one out of the four multiple images of the clump is visible in the data, but its detection is possible thanks to extended lensing constraints, color information and the high resolution of JWST/NIRCam.
\end{itemize}

%We also discuss extensions of our analysis to improve further lensing constraints on the structure of cluster members. In particular, the surface brightness of the deflectors in \elanz is more complex than in previous HST data, which would require to go beyond the commonly-used parametric models (concentric Sérsic profiles or similar). If such models are obtained, they could provide interesting constraints on the baryonic component of these cluster members, and enable the separation from their dark matter halo. In addition, the JWST/NIRCam data of \elanz likely provide enough constraints to go beyond single power-law elliptical mass density profiles. More flexible models may allow to detect signs of high-order multipoles \citep[which so far have required milliarcsecond resolution,][]{Powell2022} and provide constraints on both the inner and outer density profiles of the deflectors \citep[see e.g.,][]{Granata2023}. We defer the use of more elaborate lens models to future works.

Photometric searches have already, and will continue to, provide new galaxy-scale lenses with prominent arcs like \elanz in cluster environments \citep[e.g.,][]{Jaelani2020,Garvin2022}. Of the order of $10^5$ new galaxy clusters are expected to be discovered from deep wide field surveys such as the one undertaken with the Euclid space telescope \citep[e.g.,][]{Euclid_Adam2019,Natarajan2024_review}, with likely similar numbers from the future Legacy Survey of Space and Time at the Rubin observatory. Such a large increase of known clusters will inevitably increase the number of galaxy-scale lensing events, which have so far been more numerous than predicted by \lcdm \citep{Meneghetti2022,Meneghetti2023}.

In the past, very few studies jointly used families of multiple images in galaxy group or cluster environments together with extended source modeling to improve constraints on the distribution mass and further probe the structure of galaxies \citep[][]{Monna2015,Wang2022}. Other studies have used state-of-the-art source modeling techniques to assess the accuracy of cluster-scale lens models and visualize the morphology of background galaxies \citep[e.g.,][]{Grillo2016,Grillo2020}, which is a step towards this goal. We have shown in our work that recent advances in forward modeling and inference techniques, boosted by GPU parallelization, enable pixel-level modeling of a large number of data pixels while maintaining low computation times. Therefore, it is within reach to expend strong lensing analyses of galaxy groups and clusters by incorporating extended arcs modeling in the near future.

%Currently, of the order of $10^4 - 10^5$ galaxy clusters are known, depending on the selection method: over $130\,000$ clusters were detected in the Sloan Digital Sky Survey \citep[][]{Wen2012}, $\lesssim 10^4$ when cross-matched with X-ray or Sunyaev–Zeldovich (SZ) surveys \citep[for recent works, see e.g.,][]{Klein2023,Bulbul2024}.

\begin{acknowledgements}
The authors warmly thank the anonymous referee for the many useful suggestions that improved this manuscript. AG acknowledges funding and support by the Swiss National Science Foundation (SNSF). AG thanks Philipp Frank and Gordian Edenhofer for fruitful discussions and help with \nifty, as well as Claudio Grillo for his thoughts on the results. AG also thanks Patrick Kamieneski for providing additional details on their lens model of \elanz, as well as Rogier Windhorst, Jake Summers, and Anton Koekemoer from the PEARLS team for providing additional details on reduced JWST data products. AG thanks the Max Planck Computing and Data Facility for support. JK and SHS acknowledge funding by the Deutsche Forschungsgemeinschaft (DFG, German Research Foundation) under Germany´s Excellence Strategy – EXC 2094 – 390783311. JR acknowledges financial support by the German Federal Ministry of Education and Research (BMBF) under grant 05A20W01 (Verbundprojekt D-MeerKAT). JK also acknowledges funding from the European Research Council (ERC) under the European Union’s Horizon 2020 research and innovation programme (Grant agreement No. 101071643). SHS thanks the Max Planck Society for support through the Max Planck Fellowship. This research has also made use of \textsc{SciPy} \citep{Virtanen2020scipy}, \textsc{NumPy} \citep{Oliphant2006numpy,VanDerWalt2011numpy}, \textsc{Matplotlib} \citep{Hunter2007matplotlib}, \textsc{Astropy} \citep{astropy2013,astropy2018}, \textsc{Jupyter} \citep{Kluyver2016jupyter} and \textsc{GetDist} \citep{Lewis2019getdist}.
\end{acknowledgements}

%-------------------------------------------------------------------

\bibliographystyle{aa}
\bibliography{biblio}

%-------------------------------------------------------------------

\begin{appendix}

\section{Weighting used in color composite images \label{app:sec:color_composites}}

In Figs.~\ref{fig:model_color} and \ref{fig:bestfit_lens_model_summary}, we constructed color composite images of JWST and HST datasets. The arbitrary weighting of individual bands, chosen for visualization purposes only, is the following:
\begin{itemize}
\setlength\itemsep{0.8em}
    \item JWST/NIRCam: ${\rm red}=2\times {\rm F444W},\, {\rm green}=1\times {\rm F356W}+1\times {\rm F410M},\, {\rm blue}=3\times {\rm F150W}+2\times {\rm F200W}+1\times {\rm F277W}$;
    \item HST/ACS+WFC3: ${\rm red}=4.2\times {\rm F160W},\,{\rm green}=1\times {\rm F140W}+1\times {\rm F125W}+1\times {\rm F105W},\, {\rm blue}=8\times {\rm F775W}+2\times {\rm F625W}+2\times {\rm F606W}+1\times {\rm F435W}$. 
\end{itemize}
The \citet{Lupton2004} scheme is then used to create the final color images.

\section{Uncertainty about the source redshift \label{app:sec:alt_src_redshift}}

As mentioned in Sect.~\ref{ssec:redshift_vel_disp}, we assume a source redshift of $\zs=2.291$ throughout this work. However, \citet{Kamieneski2023} also indicate that an alternative redshift value, although less likely, can be $\zs=3.388$ if the emission line detected in ALMA data corresponds to the CO(4-3) transition instead of CO(2-1). Here we discuss what would be the changes implicated by this alternative redshift.

Assuming a different source redshift impacts the Einstein radii estimates obtained from stellar velocity dispersion measurements (through the angular diameter distance \Ds, see Eq.~\ref{eq:veldisp2thetaE}). Assuming $\zs=3.388$, and a realistic contribution from the cluster to the total potential would increase the Einstein radii to $\theta_{\rm E, L1}=1\farcs25\pm0\farcs05$ and $\theta_{\rm E, L3}=0\farcs58\pm0\farcs06$ for L1 and L3, respectively. Compared to the values from Table~\ref{tab:props_from_muse}, this corresponds to an increase of $\sim 24\%$ for L3. The disagreement between the lensing and stellar kinematics estimates of $\theta_{\rm E, L3}$ discussed in Sect.~\ref{ssec:mass_constraints} would increase to $3.7\sigma$.

The second change caused by a different source redshift relates to the physical sizes of the reconstructed lensed features in source plane. If one assumes $\zs=3.388$, the physical size of the clump detected in Sect.~\ref{ssec:clump_detection} and its distance to the source centroid would reduce to $\sim 360$ pc and $\sim1.1$ kpc, respectively. In other words, the alternative source redshift would imply that the clump is smaller and closer to the galaxy center.

The last consequence of the uncertain source redshift is the rest-frame wavelength quoted in Fig.~\ref{app:fig:source_compact_region_seds}, which shows the SED of an unresolved source feature. Since we do not specifically use the SED, this has no impact on our results.

\section{Complementary results and discussion \label{app:sec:comp_results}}

\subsection{Surface brightness properties of the deflectors \label{app:ssec:lens_light}}

\begin{table*}
    \caption{Summary of photometric properties measured from the JWST data.}
    \label{app:tab:lens_light_props}
    \renewcommand{\arraystretch}{1.4}
    \centering
    \begin{tabular}{ccccccc}
    \hline\hline
    Deflector & \multicolumn{2}{c}{L1} & \multicolumn{2}{c}{L2} & \multicolumn{2}{c}{L3} \\ 
    JWST filter & $\theta_{\rm eff}$ $[\,''\,]$ & $m_{\rm AB}$ [mag] & $\theta_{\rm eff}$ $[\,''\,]$ & $m_{\rm AB}$ [mag] & $\theta_{\rm eff}$ $[\,''\,]$ & $m_{\rm AB}$ [mag] \\ 
    \hline
    F150W & $0.57\pm0.02$ & $18.95\pm0.05$ & $0.39\pm0.03$ & $20.68\pm0.05$ & $0.16\pm0.02$ & $20.47\pm0.05$ \\ 
    F200W & $0.61\pm0.01$ & $18.53\pm0.05$ & $0.36\pm0.02$ & $20.25\pm0.05$ & $0.15\pm0.01$ & $20.07\pm0.05$ \\ 
    F277W & $0.62\pm0.01$ & $18.32\pm0.05$ & $0.30\pm0.02$ & $20.18\pm0.05$ & $0.16\pm0.01$ & $19.780\pm0.05$ \\ 
    F356W & $0.54\pm0.01$ & $18.32\pm0.05$ & $0.29\pm0.01$ & $20.10\pm0.05$ & $0.15\pm0.02$ & $19.81\pm0.05$ \\ 
    F410M & $0.54\pm0.01$ & $18.45\pm0.05$ & $0.26\pm0.01$ & $20.13\pm0.05$ & $0.15\pm0.01$ & $19.10\pm0.05$ \\ 
    F444W & $0.60\pm0.01$ & $18.63\pm0.05$ & $0.30\pm0.01$ & $20.25\pm0.05$ & $0.15\pm0.01$ & $20.17\pm0.05$ \\ 
    \hline
    \end{tabular}
    \tablefoot{Quoted uncertainties only include statistical errors, hence are likely underestimated given the quality of fit to the imaging data. Uncertainties on the reported AB magnitudes have been artificially inflated to a more conservative $50$ mmag.}
\end{table*}

\begin{figure*}
    \centering
    \includegraphics[width=\linewidth]{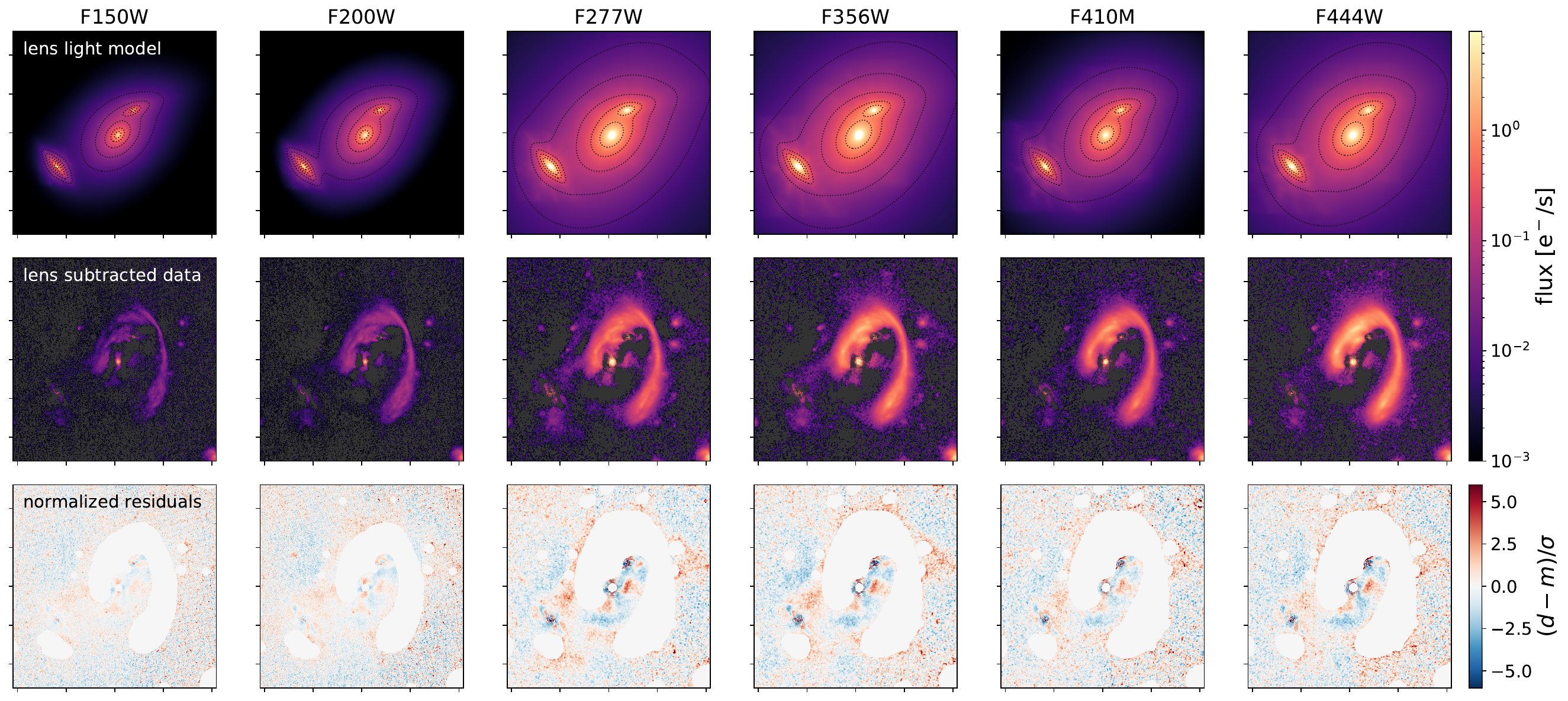}
    \caption{Best-fit models of the deflectors surface brightness of \elanz, for each of the six JWST filters considered in this work. \textit{First row}: model of the surface brightness of L1, L2 (two concentric Sérsic profiles) and L3 deflectors (three concentric profiles). \textit{Middle row}: data after subtraction of the best-fit model shown in first row. Negative pixels are displayed in dark gray in these panels. \textit{Bottom row}: normalized residuals maps corresponding to the best-fit models. White areas indicate pixels that were masked out (lensed arcs and nearby non-modeled objects) during the fit of the lens light distribution.}
    \label{app:fig:lens_light_model}
\end{figure*}

We show in Fig.~\ref{app:fig:lens_light_model} the best-fit models for the deflectors surface brightness, and measurements of effective radius and aperture magnitudes are indicated in Table~\ref{app:tab:lens_light_props}. We obtain a first-order approximation of model parameters uncertainties using the Fisher information matrix, by inverting the Hessian matrix of the loss function evaluated at the best-fit position. However, depending on the application, we caution on using these uncertainties for galaxies L1 and L2, as their surface brightness is significantly more complex than the smooth light profiles we used.

\subsection{Full joint posterior distribution of mass model parameters \label{app:ssec:lens_mass_full_post}}

\begin{figure*}
    \centering
    \includegraphics[width=\linewidth]{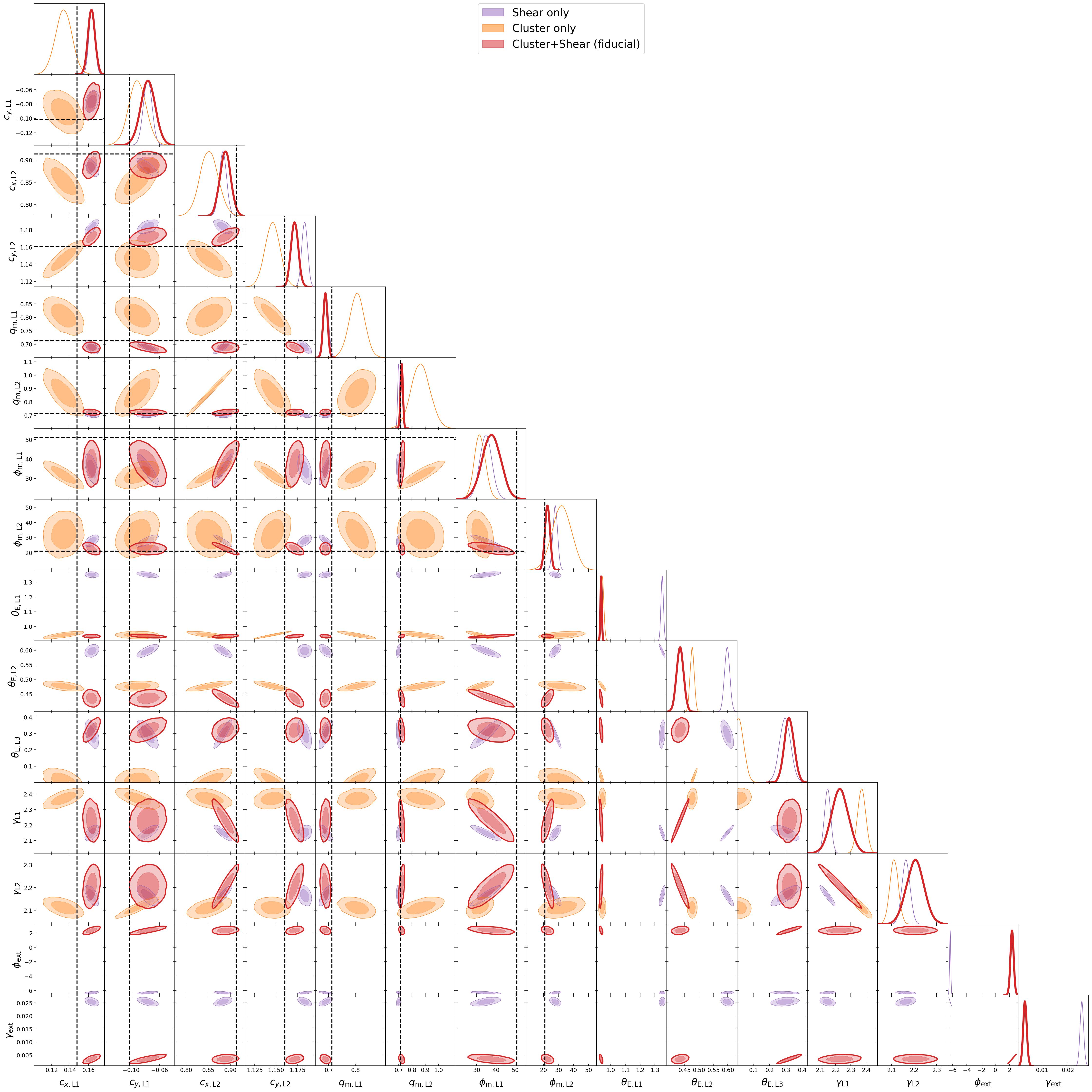}
    \caption{Full version of Fig.~\ref{fig:lens_param_corner_main}, showing the joint posterior distribution of a subset of lens mass parameters, for the 3 main model families tested in this work. The first one includes only external shear in addition of the three main deflectors, the second one replaces external shear with the \elgor cluster deflection field, and the last one includes a combination of the two. Dashed lines indicate the mean position (over the six filters), position angles and axis ratios of the deflectors light, to compare with the values inferred from lensing.}
    \label{app:fig:lens_param_corner_full}
\end{figure*}

In Sect.~\ref{ssec:mass_constraints} we present constraints on mass model parameters given the three model families that we assumed. For completeness, we show in Fig.~\ref{app:fig:lens_param_corner_full} the full version of Fig.~\ref{fig:lens_param_corner_main}, that shows the entire set of mass model parameters that are inferred (jointly with the source model). Since our lens surface brightness informs motivated priors for the centroids, axis ratio and position angle of the EPL profiles, we also indicate with dashed lines the best-fit values from our light models for L1 and L2. This completes Fig.~\ref{fig:lens_ellipses}, which illustrates the offset between the centroid of the lens and mass distributions of these elliptical galaxies.

\subsection{Robustness of the constraints \label{app:ssec:systematics}}

\begin{figure}
    \centering
    \includegraphics[width=0.7\linewidth]{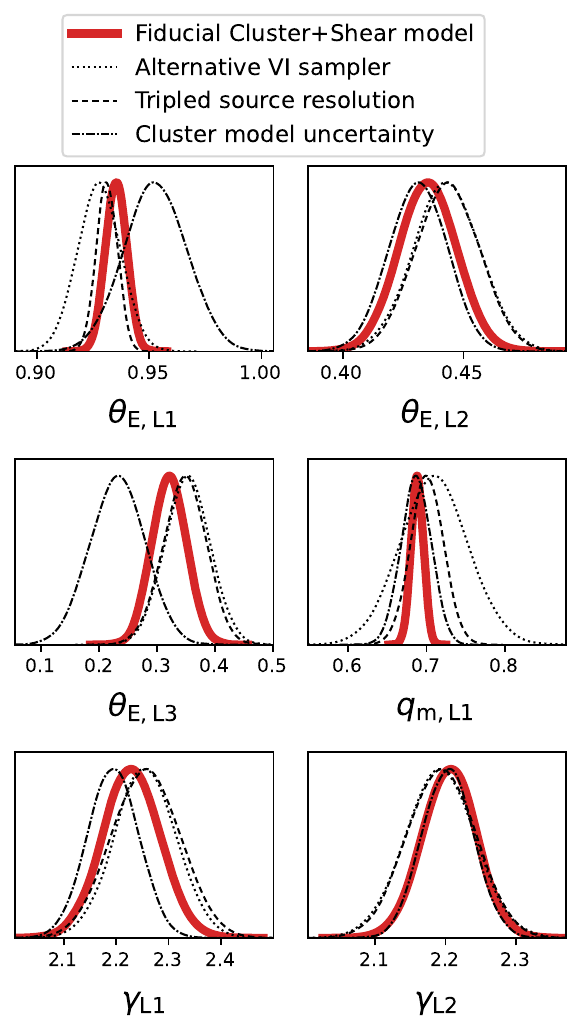}
    \caption{Systematic checks to test the robustness of the model. Three additional posterior distributions are shown, associated to different modeling and sampling choices.}
    \label{app:fig:posterior_systematic_checks}
\end{figure}

As detailed in Sect.~\ref{ssec:lens_modeling}, we already include in the marginalized posterior distributions of Fig.~\ref{fig:lens_param_corner_main} several of model variations within a given model family, including changing the initial parameters values. Here we investigate further the robustness of our inference and look for possible systematic errors due to specific modeling choices.

We vary the overall source grid resolution, namely in the number of pixels along each spatial dimension. As mentioned in Sect.~\ref{ssec:modeling_seq}, we selected the fiducial number of source pixels $n_{\rm src}=100$ from a series of preliminary models, and noticed that the $\chi^2_\nu$ value does not significantly vary when increasing $n_{\rm src}$. Nevertheless, drastically increasing the source resolution (e.g., doubling or tripling it) can still provide a slightly better fit to the data---essentially capturing unresolved point-like features in the arc. However, given the much larger number of parameters (due to the multi-band field model parametrization), these models are heavily disfavored by the BIC. Here we show the inference results after setting $n_{\rm src}=300$, namely tripling our fiducial source resolution. The resulting posteriors are shown as dotted lines in Fig.~\ref{app:fig:posterior_systematic_checks} for a subset of mass model parameters.

We also investigate whether the choice of variational inference algorithm impact the estimation of the posterior distribution. Given the large number of parameters, it may be expected to have different algorithms converging to difference locations in parameter space, potentially providing different uncertainty estimations. We do so by replacing the sampler type \texttt{'geometric'} (\textsc{geoVI}) by \texttt{'altmetric'} within \nifty. The main difference between \texttt{'geometric'} and  \texttt{'altmetric'} methods reside in the resampling step performed at each iteration. The resulting posteriors are indicated in Fig.~\ref{app:fig:posterior_systematic_checks} with dashed lines.

Finally, we investigate the impact of the uncertainty on the cluster-scale mass distribution. We randomly draw 20 posterior samples from the cluster model of \citet{Caminha2023} and run our inference pipeline using each the corresponding deflection field samples (with identical initial position in the parameter space). The resulting posteriors, marginalized over the 20 cluster deflection field posterior samples, is indicated in Fig.~\ref{app:fig:posterior_systematic_checks} with dotted-dashed lines. We note that among these cluster-scale posterior samples, some are leading to sensibly better fit to the imaging data, compared to the best-fit model we use throughout this work. Therefore, \elanz strong lensing features seem to provide extra constraints also on the large scale component (i.e., the cluster-scale dark matter halo) of the cluster.

From all the additional models we run, we find that all posterior distributions remain statistically compatible with those of our fiducial model. Therefore, we conclude that, among the many modeling choices we have explored, we do not find a significant source of systematic error in our lens model of \elanz.

\subsection{Fifth and central images in strong lenses \label{app:ssec:central_images}}

As shown in Sect.~\ref{ssec:image_families}, we confirm the five-image lensing configuration of \elanz. This is a rather rare situation: in only a few galaxy-scale strong lenses with a single or two main deflectors has been discovered a fifth extended, non-demagnified image. The vast majority of fifth images fall in the category of unresolved demagnified central images, which appear close to the deflector centroid and are often only detected at radio wavelengths thanks to weaker contamination from the lens light \citep[e.g.,][]{Gavazzi2003,Tamura2015,Wong2015,Collett2017}. The appearance of extended source features in the vicinity of L1 and L3 centroids may provide constraints on the inner part of the mass density profile of the deflectors (at the cost of larger blending between the light of the various components). A comparable configuration in the literature is the Cosmic Horseshoe strong lens, for which \citet{Schuldt2019} used the image of a second lensed source appearing close to the lens galaxy to constrain the inner dark matter profile of that galaxy. However, as it is the case for the Cosmic Horseshoe, the fifth image often corresponds to a ``radial'' image (as opposed to much more common ``tangential'' images). Coming back to \elanz, the fifth of the source remains a tangential one, thus making it a very peculiar lensing configuration. Although not entirely comparable, strong lens systems in which a low-mass satellite galaxy coincides with a strongly lensed arc and locally creates additional strong lensing features can give rise to additional tangential images \citep[see systems analysed in][]{SuyuHalkola2010,Vegetti2010_theclone}.

\section{Spectral energy distribution of a source clump \label{app:sec:clump_sed}}

\begin{figure}
    \centering
    \includegraphics[width=0.8\linewidth]{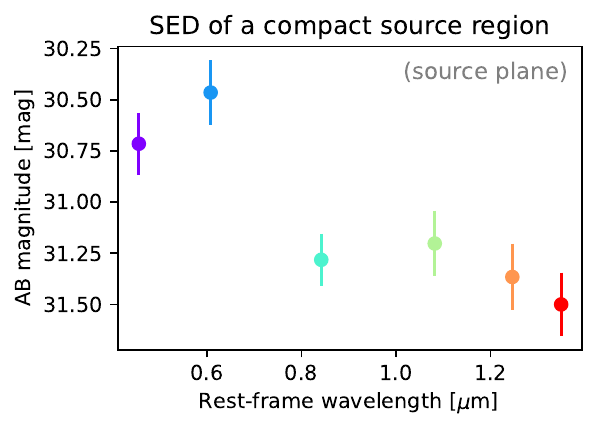}
    \caption{Spectral energy distribution of image 2b shown in the top right panel of Fig.~\ref{fig:multiple_images}, from the six JWST/NIRCam filters considered in this work. The $x$ axis shows rest-frame wavelength assuming a source redshift of $\zs=2.291$ and the $y$ axis shows de-magnified AB magnitudes, integrated within a circular aperture of radius $0\farcs2$, after subtraction of both the lens and source light models.}
    \label{app:fig:source_compact_region_seds}
\end{figure}

We show in Fig.~\ref{app:fig:source_compact_region_seds} the spectral energy distribution (SED) of one of the point-like features in the western part of the arc (image 2b in Fig.~\ref{fig:multiple_images}). We subtract the lens light, the arc light and integrate the observed flux within a circular aperture with radius $0\farcs2$. Since our source model does not capture extremely compact (likely unresolved) features like image 2b and other counter-images, it can be subtracted from the data to perform reasonably good aperture photometry.

\end{appendix}

\typeout{get arXiv to do 4 passes: Label(s) may have changed. Rerun}

\end{document}